\documentclass[12pt, draftclsnofoot, onecolumn]{IEEEtran}

\usepackage{subfig}
\usepackage{float}
\usepackage{amssymb}

\newtheorem{theorem}{Theorem}
\newtheorem{lemma}{Lemma}
\newtheorem{example}{Example}
\usepackage{amsmath}
\usepackage{graphicx}
\usepackage{comment}
\usepackage{cite}

\title{\LARGE \bf
Novel Time Asynchronous NOMA schemes for Downlink Transmissions 
}

\author{Mehdi Ganji$^{1}$ and Hamid Jafarkhani$^{2}$
\thanks{}
\thanks{$^{2}$H. Jafarkhani is with Faculty of Electrical Engineering, University of California, Irvine}%
\thanks{$^{1}$M. Ganji is with the Department of Electrical Engineering, University of California, Irvine}%
}

\begin{document}

\author{Mehdi~Ganji,~\IEEEmembership{Student Member,~IEEE,}
Hamid~Jafarkhani,~\IEEEmembership{Fellow,~IEEE} \thanks{M. Ganji and H. Jafarkhani are with the Center for Pervasive Communications and Computing, University of California, Irvine, CA, 92697 USA (e-mail:
\{mganji, hamidj\}@uci.edu). This work was supported in part by the NSF Award CCF-1526780.}}

\maketitle

\begin{abstract}

In this work, we investigate the effect of time asynchrony in non-orthogonal multiple access (NOMA) schemes for downlink transmissions. First, we analyze the benefit of adding intentional timing offsets to the conventional power domain-NOMA (P-NOMA). This method which is called Asynchronous-Power Domain-NOMA (AP-NOMA) introduces artificial symbol-offsets between packets destined for different users. It reduces the mutual interference which results in enlarging the achievable rate-region of the conventional P-NOMA. Then, we propose a precoding scheme which fully exploits the degrees of freedom provided by the time asynchrony. We call this multiple access scheme T-NOMA which provides higher degrees of freedom for users compared to the conventional P-NOMA or even the modified AP-NOMA. T-NOMA adopts a precoding at the base station and a linear preprocessing scheme at the receiving user which decomposes the broadcast channel into parallel channels circumventing the need for Successive Interference Cancellation (SIC). The numerical results show that T-NOMA outperforms AP-NOMA and both outperform the conventional P-NOMA. We also compare the maximum sum-rate and fairness provided by these methods. Moreover, the impact of pulse shape and symbol offset on the performance of AP-NOMA and T-NOMA schemes are investigated.
\end{abstract}

\section{Introduction}

For future radio access, significant gains in the system capacity/efficiency and quality of user experience are required. In particular, the multiple access approach is a key part of radio access technology \cite{dai2015non}. In \cite{kishiyama2013future,vanka2012superposition}, and the references therein, non orthogonal multiple access (NOMA) is proposed as a candidate of future radio access to partially fulfill the requirements of future networks and the possibility of downlink NOMA for 5G is currently being examined by 3GPP.
The currently prevailing approach for multiple access lies in the category of orthogonal multiple access (OMA). In 2G systems, time
division multiple access (TDMA) is adopted . In the 3G mobile communication systems such as W-CDMA and CDMA2000, direct sequence-code division multiple access (DS-CDMA) is used and the receiver is based on simple single-user detection using the Rake receiver. OMA based on orthogonal frequency division multiple access (OFDMA) or single carrier-frequency division multiple access (SC-FDMA) is used in the 4th generation mobile communication systems such as LTE and LTE-Advanced. These approaches first partition resources into orthogonal resource blocks and then assign each resource block exclusively to one user. After this, the problem is reduced to a point-to-point (P2P) communication problem and then well developed single-user encoders/decoders can be applied. The significant advantage of OMA methods is that their complexity is merely the complexity of single-user encoders/decoders. On the other hand, assigning resource blocks exclusively can be very inefficient (in terms of achievable rate-regions) and may pose a serious fairness problem among users. In contrast to OMA, NOMA allows users to utilize the same resource blocks for transmission simultaneously and therefore is potentially more efficient. In fact, when evaluated under the LTE system characteristics, NOMA demonstrates significant gains over OMA systems \cite{saito2013non,saito2013system}.

 The problem of communicating with many receivers arises in many ``downlink'' scenarios such as communication from an access point to stations in WiFi or from a base station in cellular systems. Although OMA approaches eliminates interference between transmissions, it does not in general achieve the highest possible transmission rates for a given packet error rate \cite{tse2005fundamentals}. In fact, superposition Coding (SC) is a well-known non-orthogonal scheme that achieves the capacity on a scalar Gaussian broadcast channel \cite{cover1972broadcast}. Superposition coding is a technique of simultaneously communicating information to several receivers by a single source. In other words, it allows the transmitter to send the information of multiple users at the same time and frequency. At the receiver's side successive interference cancellation (SIC) is applied which exploits the differences in signal strength among the signals of interest \cite{jafarkhani2005space}. The basic idea behind SIC is that user signals are successively decoded. In fact, superposition coding and SIC are the optimal encoder/decoder methods for degraded broadcast channels where users can be ordered in terms of the quality of the received signals \cite{cover2012elements,tse2005fundamentals}. This is of particular importance in cellular systems where the channel conditions vary significantly among users due to the near-far effect\cite{vanka2012superposition}. From an information-theoretic perspective, NOMA with a SIC is an optimal multiple access scheme from the viewpoint of the achievable multiuser capacity region, in the downlink \cite{caire2003achievable,viswanath2003sum,vishwanath2003duality,yu2004sum,weingarten2006capacity} and in the uplink \cite{wyner1994shannon}. 

Applications of NOMA in the downlink scenario have been widely studied \cite{xu2015new}. In \cite{di2015radio} and \cite{liu2015user}, various power allocation and user scheduling algorithms were proposed to improve the sum-rate of the NOMA-based multi-user system. In \cite{ding2015cooperative}, cooperative NOMA scheme was investigated to improve the spectral efficiency and transmission reliability. More recently, the study of the combination of multiple-input and multiple-output (MIMO) and NOMA has received considerable attention \cite{choi2016power}, \cite{ding2016application}. However, most of the previous research on NOMA only considered symbol-synchronous transmission. In fact, often in the literature, timing mismatch is considered as an impairment and different synchronization methods are applied to eliminate it \cite{nasir2016timing}. However, in this work, we show that time asynchrony can indeed be beneficial. By using proper transmission and receiver design, time asynchrony can decrease interference and also provide additional degrees of freedom which can be exploited to improve the performance.

The usefulness of timing offset, or time asynchrony, have been studied in the literature. For example, the results in \cite{verdu} show that time asynchrony can increase the capacity region in multiple-access  channels. Also, time asynchrony can improve the performance in other scenarios if the proper sampling and detection methods are used \cite{cui2017asynchronous,poorkasmaei2015asynchronous,avendi2015differential,ganji2016interference,das2011mimo,ganji2018performance,zou2018uplink}. 

In this work, we introduce two schemes called AP-NOMA and T-NOMA which use time asynchrony to improve and enlarge the downlink rate-region. We have the following specific contributions:
\begin{itemize}
\item We introduce two NOMA broadcast methods, AP-NOMA and T-NOMA which exploit the intentionally added timing offsets between the superimposed signals and improve the performance of the conventional P-NOMA
\item We analytically prove that for a wide range of pulse shaping filters including, rectangular, sinc and raised cosine, AP-NOMA decreases the inter-user interference (IUI), thus improves the overall performance. 
\item We demonstrate that T-NOMA can take advantage of limited time communication by appropriate precoding in order to provide higher degrees of freedom and hence improve the performance. 
\item We derive the achievable rate-regions for the proposed schemes for arbitrary number of users and provide numerical results.  
\item We analytically show that for the P-NOMA method to achieve the maximum sum-rate, it needs to assign all the power to the strongest user, thus violating the fairness. However, AP-NOMA and T-NOMA methods can achieve the maximum sum-rate while maintaining fairness among the users. 
\end{itemize}

The rest of the paper is organized as follows: 
In Section \ref{noma}, we provide an overview of the concepts of NOMA including superposition coding and SIC. Next, we provide some insights regarding the benefits of time asynchrony. The concept behind the proposed method AP-NOMA is the reduction in IUI by using intentional time delays, and the concept behind T-NOMA is exploiting extra degrees of freedom provided by time asynchrony which are explained in Section \ref{asynch}. Then, we present the system model and its characteristics in Section \ref{model}. We present the achievable rate-region results for the conventional P-NOMA, modified AP-NOMA and T-NOMA in Section \ref{analysis}. At the end, we provide the numerical results and the final remarks in Sections \ref{simulation} and
\ref{conclusion}, respectively. 
\section{Concepts of P-NOMA}\label{noma}

Simultaneous transmission of information from one source to several receivers has been studied under the title of broadcast channel\cite{cover1998comments}. Superposition coding at the transmitter and SIC at the receivers provide the capacity achieving performance and thus play important roles in the P-NOMA. In this section, we briefly summarize the results in the literature and explain the concepts of superposition coding and SIC method.  
Consider the Gaussian broadcast channel
\begin{flalign}
\nonumber
Y_1&=X+Z_1\\
\nonumber
Y_2&=X+Z_2
\end{flalign}
where $Z_1$ and $Z_2$ follow Gaussian distributions, i.e., $Z_1\sim N(0,N_1)$ and $Z_2\sim N(0,N_2)$, assuming that $N_2>N_1$. 
\begin{theorem}
The capacity region for the Gaussian broadcast channel, with signal power constraint $P$, is given by:
\begin{flalign}
\nonumber
R_1&\leq \frac{1}{2}\log\left(1+\frac{\alpha P}{N_1}\right)\\
\nonumber
R_2&\leq \frac{1}{2}\log\left(1+\frac{(1-\alpha) P}{\alpha P+N_2}\right)
\end{flalign}
This region is achieved by the superposition coding and SIC schemes described in \cite{cover1972broadcast}.

\end{theorem}

\begin{figure}[h]
    \centering
    \includegraphics[width=2in]{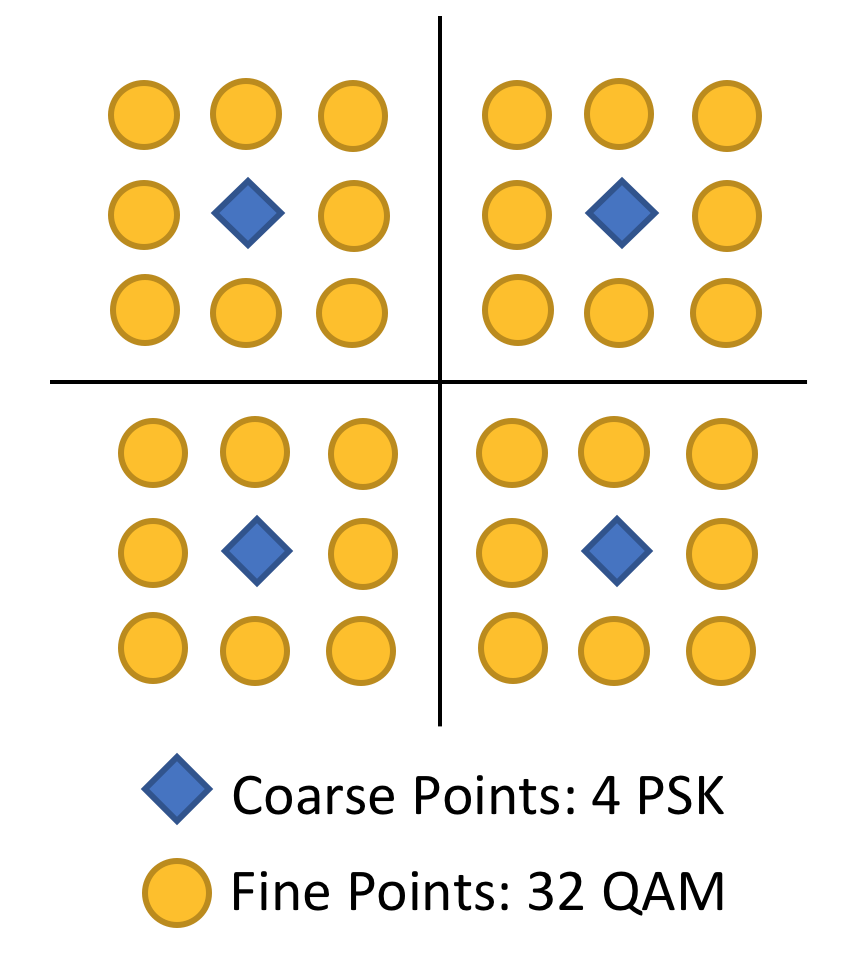}
    \caption{An Example of Superposition Coding}
    \label{constel}
\end{figure}

To encode the messages, the transmitter generates two codebooks, one with power $\alpha P$ at rate $R_1$, and another codebook with power $(1-\alpha)P$ at rate $R_2$, where $R_1$ and $R_2$ lie in the capacity region above. Then, to send an index $w_1\in \{1,2,\cdots,2^{nR_1}\}$ and $w_2 \in \{1,2,\cdots,2^{nR_2}\}$ to $Y_1$ and $Y_2$, respectively, the transmitter takes the codeword $X(w_1)$ from the first codebook and codeword $X(w_2)$ from the second codebook and computes the sum. The transmitter sends the sum over the channel \cite{cover2012elements}. An example of superposition Coding using 4-PSK and 8-QAM constellations for two users are shown in Fig. \ref{constel}. The input bits of the user with weaker channel are modulated with 4-PSK modulation as shown by the coarse points in Fig. \ref{constel} and the input bits for the user with stronger channel are modulated with 8-QAM modulation and sum of the modulated symbols results in a 32-QAM constellation shown by the fine points in Fig. \ref{constel}. 

\begin{figure}[h]
\centering
\subfloat[Decoding at the weak user/first step at the strong user]{\includegraphics[width=2in]{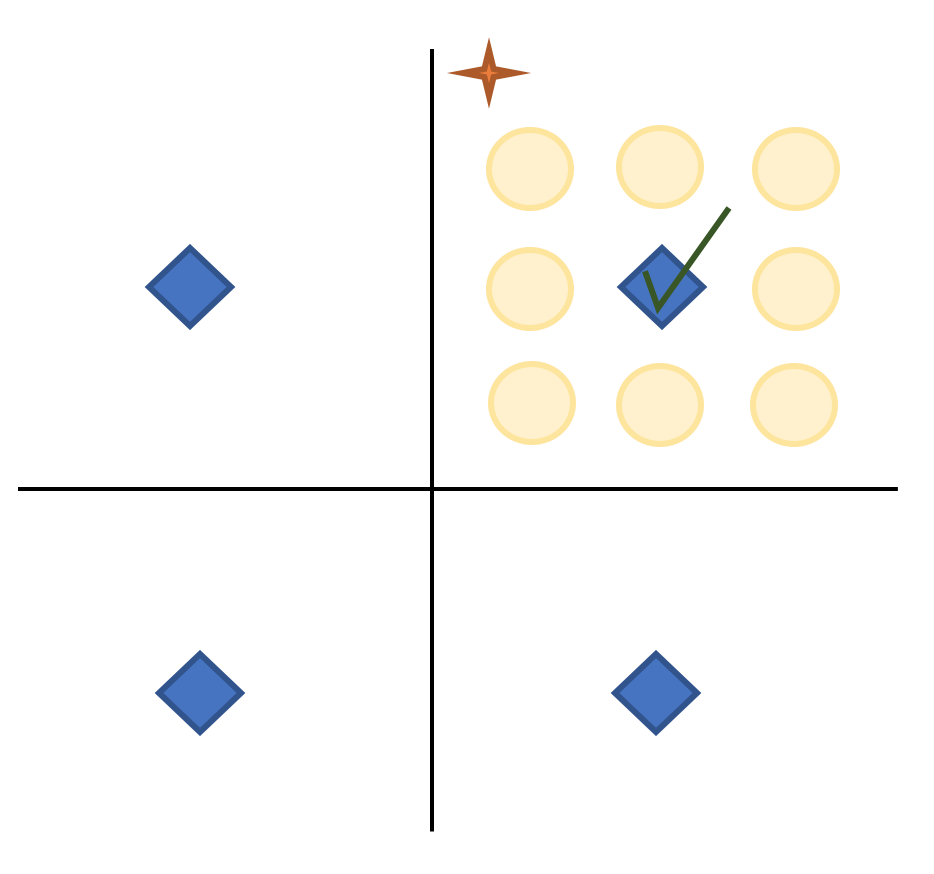}}
\hfil
\subfloat[Second step of decoding at the strong user]{\includegraphics[width=2in]{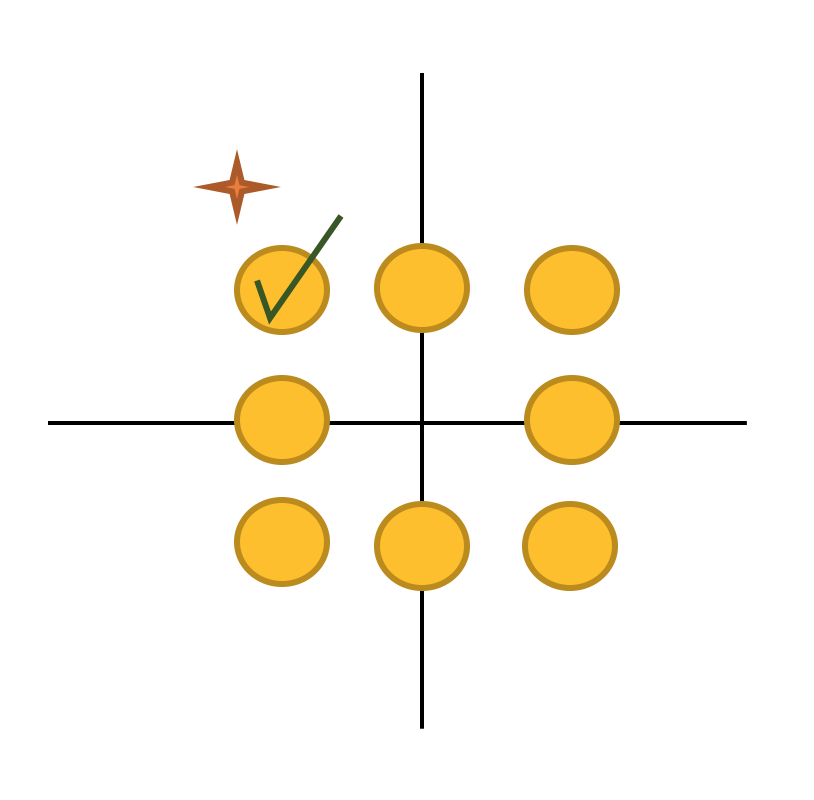}
\label{decode2}}
\caption{An example of SIC decoding}
\label{decode}
\end{figure}

The receivers must now decode the messages. The weak receiver, $Y_2$, merely looks through the second codebook to find the closest codeword to the received vector $Y_2$. His effective signal-to-noise ratio is $\frac{(1-\alpha)P}{\alpha P +N_2}$, since $Y_1$'s message acts as noise to $Y_2$. The strong receiver, $Y_1$, first decodes $Y_2$'s codeword, which he can accomplish because of his lower noise $N_1$. After subtracting this codeword $\hat X_2$ from $Y_1$, the receiver looks for the codeword in the first codebook closest to $Y_1-\hat X_2$. The resulting probability of error can be made as low as desired. A nice dividend of optimal encoding for degraded broadcast channels is that the strong receiver $Y_1$ always knows the message intended for receiver $Y_2$ in addition to its own message \cite{cover2012elements}. Fig. \ref{decode} presents the technique for decoding the superposed signal (Fig. \ref{constel}) at the receiving side. As shown in Fig. \ref{decode}, the weak receiver only decodes the coarse points by mapping the received signal to the nearest point in the corresponding constellation (4-PSK). The stronger user is also able to decode the coarse points and after subtracting the decoded symbol from the received signal, the resulting signal is decoded using the corresponding constellation (8-QAM) as shown in Fig. \ref{decode2}.

\section{Motivations Behind Asynchronous Transmission}\label{asynch}
It is shown in the literature that time asynchrony which is the intrinsic feature of most of the wireless networks, not only is not disruptive, but also can be beneficial if the proper processing is employed \cite{cui2017asynchronous,poorkasmaei2015asynchronous,avendi2015differential,ganji2016interference,das2011mimo,ganji2018performance,zou2018uplink}. We introduce two NOMA schemes enjoying the benefits of time asynchrony, i.e., AP-NOMA and T-NOMA. The details of these two methods will be elaborated later, but before that, let us briefly express the intuition and motivations behind each of the mentioned methods.  

\subsection{Motivation Behind AP-NOMA}
\begin{figure}[!h]
\centering
\subfloat[Synchronous Reception]{%
\includegraphics[width=4in]{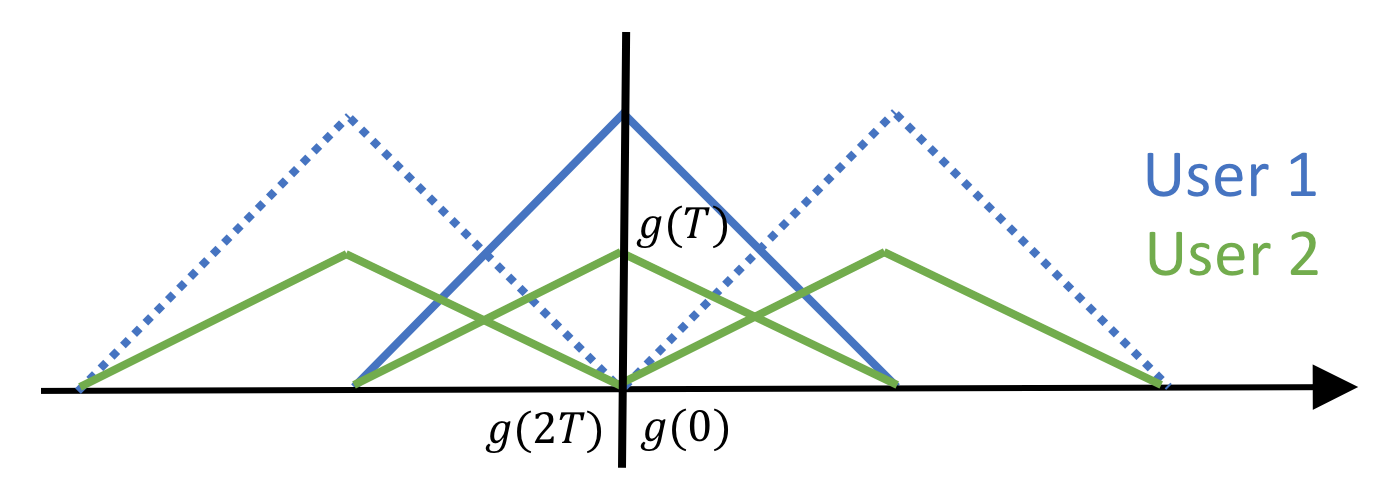}
\label{sync}}
\hfil
\subfloat[Asynchronous Reception]{%
\includegraphics[width=4in]{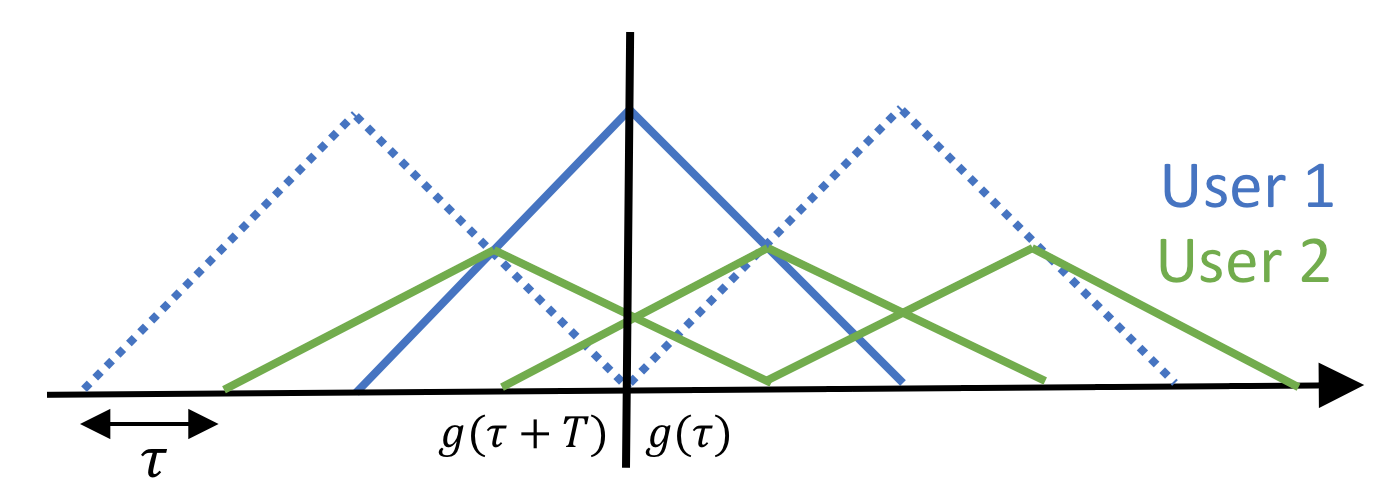}
\label{async}}
\caption{Illustration of IUI for two cases of synchronous and asynchronous reception}
\label{comp}
\end{figure}

In conventional P-NOMA, the underlying assumption is the reception with perfect synchronization. In the perfect synchronous scenario, the peak points of all users are aligned, however, by adding intentional time delays to each user, the peak points drift apart. The comparison between synchronous and asynchronous reception for two users is shown in Fig. \ref{comp}. 

Denoting the overall pulse shape, including the transmit pulse shape and the matched filter, as $g(t)$, the interuser interference (IUI) power from interfering user can be calculated as:
\begin{flalign}
IUI(\tau)=\sum_{i=-\infty}^{\infty}{|g(\tau+iT)|^2}
\end{flalign}
where $\tau$ is the time delay and $T$ is the symbol interval. In the next lemma, we show that adding intentional time delay will decrease the IUI power.
\begin{lemma}
For any pulse shape, denoted as g(t), with real spectrum in frequency domain (i.e., real and even in the time domain), we will have
\begin{flalign}
IUI(\tau)\leq IUI(0) \ \ \ 
\end{flalign}
\end{lemma}
\begin{IEEEproof}
The proof is presented in Appendix \ref{appendix1}.
\end{IEEEproof}
Note that the conditions mentioned above encompasses almost all the the pulse shapes in the literature like rectangular and sinc pulse shapes and even practical pulse shapes like the raised cosine pulse shape. Thus, adding time asynchrony can decrease the IUI which is the main degradation in NOMA schemes. However, the benefits of time asynchrony is not limited to decreasing IUI power. In fact, as we will explain later, adding time asynchrony provides additional degrees of freedom which can be exploited to serve more users over the same time and frequency resources. 

\subsection{Motivation Behind T-NOMA}
Here, we use Hilbert space formulation to show the potential of asynchronous transmission in providing additional degrees of freedom. Hilbert space generalizes the Euclidean space of real numbers to finite energy signals. Each finite energy signal can be represented by a vector in the Hilbert space with each coordinate given by an inner product with the corresponding orthonormal basis functions. In more details, any finite energy signal like $x(t)$ can be written as a linear combination of the orthonormal basis functions as:
\begin{flalign}
x(t)=\sum_{n\in \Gamma}x[n]p_n(t)
\end{flalign}
where $p_n(t)$ is an orthonormal basis function, i.e, $\langle p_n(t).p_m(t)\rangle=\delta(n-m)$ and $x[n]$ is the corresponding coefficient in the direction of the basis function $p_n(t)$ which is obtained by the following inner product.
\begin{flalign}
x[n]=\langle x(t).p_n(t)\rangle
\end{flalign}
If we further constrain our finite energy signals to strictly band-limited ones, then the Hilbert Space is called the Paley-Wiener space. The Nyquist sampling theorem states that any signal in Paley-Wiener space whose Fourier transform is supported on $f \in (-W \ W)$ can be written as the linear combination of some sinc pulses, i.e, 
\begin{flalign}\label{sinc}
x(t)=\sum_{n=-\infty}^{\infty}x[n]\left(\sqrt{2W}sinc(2W(t-(n-1)T))\right)
\end{flalign}
where $T$ is the Nyquist interval, i.e., $T=\frac{1}{2W}$ \cite{debnath2005hilbert}. Due to completeness of the of sinc pulses, all band-limited signals, even sinc pulses that do not lie at integer multiples of $T$, e.g., their shifted version, i.e., $sinc(2W(t-(n-1)T-\tau))$, still lie completely in the Paley-Wiener space. 

Assume that, in Eq. (\ref{sinc}), $x[n]$ is the transmitted symbol modulated on the sinc pulse $p_n(t)=\sqrt{2W}sinc(2W(t-(n-1)T))$. In practice, pulses spanning an unlimited time domain are not feasible, hence, they are usually truncated within a desired interval. Assume that the transmission interval is truncated into $NT$ seconds, then we are capable of transmitting approximately $2WNT$ symbols. In other words, $2WNT$ dimensions is used in the case of finite-time transmissions \cite{gallager2008principles}. However, due to the truncation, the finite set of sinc pulses, i.e.:
\begin{flalign}
S\equiv
\begin{Bmatrix}
p_n(t)=\sqrt{2W}sinc(2W(t-(n-1)T))\\
n=1,\cdots, 2WNT
\end{Bmatrix}
\end{flalign}
is not complete anymore and does not span the whole signal space. Therefore, we can insert additional pulses to exploit more signaling dimensions which leads to higher data throughput \cite{kim2013faster}. 

For example, defining $b_{2WNT}=\sqrt{2W}sinc(2W(t-\tau))$ and applying the well-known Gram-Schmidt orthogonalization process, provide us an orthonormal basis function with size $2WNT+1$. The newly formed basis function exploits an additional signaling dimension. We provide a numerical example next.

\begin{example}
Assume that $W=0.5,N=5,T=1$, then $p_n(t)=sinc(t-(n-1))$ for $n=1,\cdots, 5$  with truncation length of $5T$. The Gram matrix of the aforementioned set is equal to:
\begin{flalign}
\boldsymbol{G_S}=
\begin{pmatrix}
 0.959& 0.052 & -0.084 & 0.057 & -0.013\\ 
 0.052&0.959  & 0.052 & -0.084 & 0.057\\ 
 -0.084& 0.052 &0.959  & 0.052 & -0.084\\ 
 0.057& -0.084 & 0.052 & 0.959 & 0.052 \\ 
 -0.013 & 0.057 & -0.084  &  0.052 & 0.959
\end{pmatrix}
\end{flalign}
If the pulse shapes were strictly band-limited, i.e., unlimited time support, matrix $\boldsymbol{G_S}$ would be the identity matrix. In addition, with unlimited time support, any other function like $p_6(t)=sinc(t-0.5)$ can be written as $\sum_{n=-\infty}^{\infty}{a_n sinc(t-n)}$  where $a_n=sinc(n-0.5)$. On the other hand, in a time-limited scenario, $p_6(t)$ cannot be written as the weighted sum of truncated sinc functions. Thus, performing the Gram-Schmidt process, we can get the following orthonormal set:
\begin{flalign}
\nonumber
\begin{Bmatrix}
p'_1(t)=sinc(t), e_1(t)=\frac{p'_1(t)}{|p'_1(t)|}\\
p'_2(t)=p_2(t)-0.053e_1(t), e_2(t)=\frac{p'_2(t)}{|p'_2(t)|}\\
p'_3(t)=p_3(t)+0.086e_1(t)-0.058e_2(t), e_3(t)=\frac{p'_3(t)}{|p'_3(t)|}\\
p'_4(t)=p_4(t)-0.059e_1(t)+0.089e_2(t)-0.064e_3(t), e_4(t)=\frac{p'_4(t)}{|p'_4(t)|}\\
p'_5(t)=p_5(t)+0.014e_1(t)-0.06e_2(t)+0.091e_3(t)-0.066e_4(t), e_5(t)=\frac{p'_5(t)}{|p'_5(t)|}\\
p'_6(t)=p_6(t)-0.647e_1(t)-0.612e_2(t)+0.191e_3(t)-0.099e_4(t)+0.06e_5(t), e_6(t)=\frac{p'_6(t)}{|p'_6(t)|}
\end{Bmatrix}
\end{flalign}
Now we have a new set with six elements, i.e., $S'\equiv\{e_n(t), n=1,\cdots,6\}$, for which the Gram matrix is the identity matrix. 
\end{example}

We can continue this procedure and take advantage of the rest of the available signaling dimensions. In fact, it is shown in the literature that the available degree of freedom in a time-limited channel is unbounded \cite{barman2009capacity}. We will use asynchronous transmission to fully exploit the available degrees of freedom in a NOMA framework. In the next section, the general system model and its characteristics for the asynchronous transmission is explained. 
\section{System Model}\label{model}
After performing coding and modulation, the modulated symbols intended for each user, namely, User $k$,  are shaped with appropriate waveforms suited to the communication channel, particularly its bandwidth ($W$). We denote the block length by $N$, then the intended signal for User k will be:
\begin{flalign}
x_k(t)=\sum_{n=1}^{N}\sqrt{P_{kn}}{x_k[n]p(t-(n-1)T)}
\end{flalign}
where $p(t)$ is the pulse shape, e.g., root raised cosine, which is truncated and its length is denoted by $T_p$ and $T$ is the symbol interval and is usually equal to $\frac{1}{2W}$. The transmit power assigned to $x_k(t)$ is denoted by $P_k$ which is calculated as:
\begin{flalign}
P_k=E\left[\int_{-\infty}^{\infty} x_k(t)x_k(t)^* dt\right]
\end{flalign}
The relation between the transmit power, $P_k$ and individual symbol magnitudes of symbols $P_{kn}$ will be examined later. The transmitted signal from BS will be the super-position of signals from all users, i.e.:
\begin{flalign}
x_{synch}(t)=\sum_{k=1}^{K}{x_k(t)}
\end{flalign}
where the transmit power, i.e., $P_{synch}=E\left[\int_{-\infty}^{\infty} x_{synch}(t)x_{synch}(t)^* dt\right]$, satisfies the total power constraint for $N$ time slots, i.e., $P_{synch}\leq NP$. To take advantage of asynchrony, assume that each sub-stream is shifted with a specific time delay $\tau_k$. Then, the transmitted signal will be:
\begin{flalign}
\label{goosfand}
x_{asynch}(t)=\sum_{k=1}^{K}{x_k(t-\tau_k)}
\end{flalign}
such that $P_{asynch}\leq NP$. By assuming flat fading and additive white Gaussian noise, the received signal at the $r$th user is described as:
\begin{flalign}
y^r(t)=h_r\sum_{k=1}^{K}{x_k(t-\tau_k)}+n^r(t)
\end{flalign}
To detect the transmitted symbols $x_k[n]$, instead of working with the continuous random process $y^r(t)$, we use a set of statistics, i.e., $Z_j=r_j(y^r(t)), j=1,\cdots, J$ that are sufficient for detection of transmitted symbols. We use the well-known factorization theorem to find the sufficient statistics.
\begin{theorem}
Let $Y_1, Y_2,\cdots,Y_n$ be random variables with joint density $f(y_1,y_2,\cdots,y_n|\theta)$. The statistics
\begin{flalign}
Z_j=r_j(Y_1,Y_2,\cdots,Y_n), j=1,\cdots, J
\end{flalign}
are jointly sufficient to estimate $\theta$ if and only if the joint density can be factored as follows:
\begin{flalign}
\nonumber
&f(y_1,y_2,\cdots,y_n|\theta)=u(y_1,y_2,\cdots,y_n).\\
&v(r_1(y_1,y_2,\cdots,y_n),\cdots,r_J(y_1,y_2,\cdots,y_n),\theta)
\end{flalign}
where $u$ and $v$ are non-negative functions \cite{mendel1995lessons}. 
\end{theorem}
Assuming the knowledge of channel coefficients, power assignments and time delays, the density of $y^r(t)$ given the transmitted symbols is calculated as follows:
\begin{flalign}
\nonumber
f(y^r(t)|\{x_k[n]\})=
\nonumber
c\exp{\left[\int_{-\infty}^{\infty}{[z(t)]^2dt}\right]}
\end{flalign}
where $z(t)=y^r(t)-h_r\sum_{k=1}^K\sum_{n=1}^{N}{\sqrt{P_{kn}}x_k[n]p(t-(n-1)T-\tau_k)}$ and $c$ is a constant value independent of the transmitted symbols. Expanding the likelihood function, we will have 
\begin{flalign}
\nonumber
f(y^r(t)|\{x_k[n]\})=u_1(y^r(t)).u_2(\boldsymbol{x}).Re\left\{h^*_r\sum_{k=1}^K\sum_{n=1}^{N}{\sqrt{P_{kn}}x^*_k[n]\int_{-\infty}^{\infty}{y^r(t)p(t-(n-1)T-\tau_k)dt}}\right\}
\end{flalign}
where $u_1(y^r(t))$ and $u_2(\boldsymbol{x})$ are only functions of the output signal and the input symbols, respectively.
Therefore, using the factorization theorem stated above, we can conclude that sufficient statistics for detecting the transmitted symbols are:
\begin{flalign}
\label{suff}
y^r_l[m]=&\int_{-\infty}^{\infty}{y^r(t)p(t-(m-1)T-\tau_l)dt}\\
\nonumber
&l=1,\cdots,K \ \ \ m=1,\cdots,N
\end{flalign}
which is similar to the matched filter for the synchronous case, although it involves $K$ times more samples, and can also be implemented using convolution followed by a sampler, i.e.,:
\begin{flalign}
y^r_l[m]=y^r(t)*p(t)|_{(m-1)T+\tau_l}
\end{flalign}
Denoting $p(t)*p(t)$ as $g(t)$ and $g((m-n)T+(\tau_l-\tau_k))$ as $g_{lk}(m-n)$, the sufficient statistics can be represented as:
\begin{flalign}
y_l^r[m]=h_r\sum_{k=1}^K\sum_{n=1}^{N}{\sqrt{P_{kn}}g_{lk}(m-n)x_k[n]}+n^r_l[m]
\end{flalign}
where $n^r_l[m]=n(t)*p(t)|_{(m-1)T+\tau_l}$. Defining $\boldsymbol{y_l^r}=\left(y^r_l[1],\cdots,y^r_l[N]\right)^T$, $\boldsymbol{P_k}=$ diag$(\sqrt{P_{k1}},\sqrt{P_{k2}},\cdots, \sqrt{P_{kN}})$, and $\boldsymbol{x_k}=\left(x_k[1],\cdots,x_k[N]\right)^T$, the input-output system representation in matrix form is as follows:
\begin{flalign}
\boldsymbol{y^r}=h_r\boldsymbol{RPx}+
\boldsymbol{n^r}
\end{flalign}
where $\boldsymbol{y^r}=\left(\boldsymbol{y^r_1}^T,\cdots,\boldsymbol{y^r_K}^T\right)^T$, $\boldsymbol{x}=\left(\boldsymbol{x_1}^T,\cdots,\boldsymbol{x_K}^T\right)^T$, $\boldsymbol{n^r}=\left(\boldsymbol{n^r_1}^T,\cdots,\boldsymbol{n^r_K}^T\right)^T$ and matrices $\boldsymbol{P}$ and $\boldsymbol{R}$ are defined as:
\begin{flalign}
\boldsymbol{P}=\left(\begin{matrix}
 \boldsymbol{P_{1}}&  \boldsymbol{0}&  \cdots& \boldsymbol{0}\\ 
 \boldsymbol{0}&  \boldsymbol{P_{2}}&  \ddots& \vdots\\ 
 \vdots& \ddots & \ddots & \boldsymbol{0} \\ 
 \boldsymbol{0}& \cdots & \boldsymbol{0} & \boldsymbol{P_{K}}
\end{matrix}\right), 
\boldsymbol{R}=\left(\begin{matrix}
 \boldsymbol{R_{11}}&  \boldsymbol{R_{12}}&  \cdots& \boldsymbol{R_{1K}}\\ 
 \boldsymbol{R_{21}}&  \boldsymbol{R_{22}}&  \cdots& \boldsymbol{R_{2K}}\\ 
 \vdots& \ddots & \ddots & \vdots\\ 
 \boldsymbol{R_{K1}}& \boldsymbol{R_{K2}} & \cdots & \boldsymbol{R_{KK}}
\end{matrix}\right)
\end{flalign}
where the elements of each sub-matrix $\boldsymbol{R_{lk}}$ is defined as:
\begin{flalign}
\boldsymbol{R_{lk}}(m,n)=g_{lk}(m-n)
\end{flalign}
Matrix $\boldsymbol{R}$ is a Hermitian matrix whose sub-blocks, i.e., $\boldsymbol{R_{lk}}$ are banded Toeplitz blocks of order $u$, where $u=\frac{T_p}{T}$. Due to additional signaling, the noise samples are not independent anymore and their covariance is equal to $\boldsymbol{R}\sigma^2_n$. Note that, in the perfect synchronous case, sub-blocks $\boldsymbol{R_{lk}}$ turn into identity matrices, i.e., $\boldsymbol{I_N}$. However, for the asynchronous case, only diagonal blocks are identity matrices, i.e., $\boldsymbol{R_{ll}}=\boldsymbol{I_N}$, and the other sub-blocks have non-zero off diagonal elements. 

The question of whether this matrix is invertible or not and how it behaves asymptotically as the block length N tends to infinity will have important consequences in the performance of the system. Hence, we will investigate this question next. 

\subsection{Properties of Matrix $\boldsymbol{R}$}
To understand the asymptotic behavior of matrix $\boldsymbol{R}$, we will utilize the Szego theorem which states that\cite{gray1972asymptotic}:
\begin{theorem}
Let $\boldsymbol{T_N} = [t_{k-j};k,j = 0,1,2,\cdots,N-1]$ be a sequence of Hermitian Toeplitz matrices whose generating functions is defined as $f(w)=\sum_{k=-\infty }^{\infty}{t_ke^{ikw}}, w \in [0,2\pi]$. Also, $\lambda_{1}\geq \lambda_{2}\geq \cdots \geq \lambda_{N}$ are the sorted eigenvalues of matrix $\boldsymbol{T_N}$. Then, for any function $F$ that is continuous on the range of $f$, we have:
\begin{flalign}
\lim_{N\rightarrow \infty}\frac{1}{N}\sum_{k=0}^{N}F(\lambda_{k})=\frac{1}{2\pi}\int_0^{2\pi}F(f(w))dw
\end{flalign}
In addition, the largest and smallest eigenvalues asymptotically converge to:
\begin{flalign}
\nonumber
\lim_{N\rightarrow \infty} \lambda_{1}=\max_{w}f(w)\\
\nonumber
\lim_{N\rightarrow \infty}\lambda_{N}=\min_{w}f(w)
\end{flalign}
\end{theorem}

The proof comes from asymptotic equivalence of sequences of Hermitian Toeplitz matrices and their corresponding circulant versions which results in asymptotic convergence of their eigenvalues. Our system model is not Toeplitz in general except for the special case of equally spaced timing offsets. However, luckily, the Szego Theorem can be extended to Toeplitz block matrices \cite{gutierrez2008asymptotically}. The generalized Szego Theorem relates the collective behavior of the eigenvalues to the generalized generating function, $\boldsymbol{R(w)}$, which is defined as:
\begin{flalign}
\boldsymbol{R(w)}=\left(\begin{matrix}
 f_{11}(w)&  f_{12}(w)&  \cdots& f_{1K}(w)\\ 
 f_{21}(w)&  f_{22}(w)&  \cdots& f_{2K}(w)\\ 
 \vdots& \ddots & \ddots & \vdots\\ 
 f_{K1}(w)& f_{K2}(w) & \cdots & f_{KK}(w)
\end{matrix}\right)
\end{flalign}
where $f_{lk}(w)$ is the generating function for the corresponding Toeplitz block $\boldsymbol{R_{lk}}$. The generalized Szego Theorem states that for any continuous function $F$ \cite{7099248}:
\begin{flalign}
\lim_{N\rightarrow \infty}\frac{1}{N}\sum_{k=1}^{NK}F[\lambda_{k}(\boldsymbol{R})]=\frac{1}{2\pi}\int_0^{2\pi}\sum_{j=1}^{K}F[\lambda_{j}(\boldsymbol{R(w)})]dw
\end{flalign}
In particular, for $F(x)=x$, 
\begin{flalign}
\lim_{N\rightarrow \infty}\frac{1}{N}\sum_{k=1}^{NK}\lambda_{k}(\boldsymbol{R})=\frac{1}{2\pi}\int_0^{2\pi}\sum_{j=1}^{K}\lambda_{j}(\boldsymbol{R(w)})dw
\end{flalign}
Moreover, the same convergence results can be obtained for the largest and smallest eigenvalues, i.e,:
\begin{flalign}
\nonumber
\lim_{N\rightarrow \infty} \lambda_{1}(\boldsymbol{R})=\max_{w}\lambda_{1}(\boldsymbol{R(w)})\\
\nonumber
\lim_{N\rightarrow \infty} \lambda_{NK}(\boldsymbol{R})=\min_{w}\lambda_{K}(\boldsymbol{R(w)})
\end{flalign}
Therefore, some of the properties of matrix $\boldsymbol{R}$ can be deducted by properties of matrix $\boldsymbol{R(w)}$ when the block length is sufficiently large. For example, it is proved in the literature that for time limited transmission, i.e., finite value of u, matrix $\boldsymbol{R(w)}$ is positive definite with bounded eigenvalues, thus, eigenvalues of matrix $\boldsymbol{R}$ are nonzero and bounded. On the other hand, when the pulse shapes are strictly band-limited, matrix $\boldsymbol{R(w)}$ is singular which results in the singularity of $\boldsymbol{R}$ \cite{TorbatianMehdi2011}. 

\subsection{Transmit Power Examination}
In this section, we analyze the transmit power for synchronous and asynchronous scenarios. We calculate $P_{asynch}$ in the next lemma and $P_{synch}$ will be found by assuming that the time delays are zero.
\begin{lemma}
The transmit power of the asynchronous transmission of $K$ superimposed sub-streams defined in Eq. (\ref{goosfand}) can be calculated as:
\begin{flalign}
\nonumber
P_{asynch}&=\sum_{i=1}^{K}\sum_{j=1}^{K}trace({\boldsymbol{R_{ij}Q_{ij}}})\\
&=trace(\boldsymbol{R}\boldsymbol{Q})
\label{haji}
\end{flalign}
where $\boldsymbol{Q_{ij}}=COV[\boldsymbol{x_i,x_j}]$ and $\boldsymbol{Q}=COV[\boldsymbol{x}]$.
\end{lemma}
\begin{IEEEproof}
The proof is presented in Appendix \ref{appendix2} 
\end{IEEEproof}
Note that the total power in Eq. (\ref{haji}) is not the same as the sum of individual powers. This is because the symbols are not independent in general and $E\left[\sum{|x_k|}^2\right]$ is not equal to $E\left[|\sum{x_k}|^2\right]$, i.e., the cross-terms will not be zero. For the synchronous transmission where the sub-blocks of matrix $\boldsymbol{R}$ are identity matrices, the transmit power will be simplified to $P_{synch}=\sum_{i=1}^{K}\sum_{j=1}^{K}trace({\boldsymbol{Q_{ij}}})$. If no precoding performed in the BS, i.e., $\boldsymbol{Q_{ij}}=\boldsymbol{P_i}^2\delta[i-j]$, then the transmit power for both synchronous and asynchronous transmission will be $\sum_{k=1}^{K}{P_k}$ where $P_k=\sum_{n=1}^{N}{P_{kn}}$. Because the channel is assumed to be fixed during the transmission of a packet, the index of $n$ can be discarded for P-NOMA and AP-NOMA schemes which use the same power for all time instants. However, for T-NOMA method which uses precoding and exploits the variation of effective channel induced by asynchrony, the values of $P_{kn}$ will be assigned accordingly. Thus, the power constraint for P-NOMA, AP-NOMA and T-NOMA methods can be stated as:
\begin{flalign}
\textnormal{P-NOMA, AP-NOMA}&:\sum_{k=1}^{K}{P_k}\leq P\\
\textnormal{T-NOMA}&:trace(\boldsymbol{\boldsymbol{R}Q}) \leq NP
\end{flalign}
Note that, from now on, $P_k$ is the transmit power assigned to the transmission of each symbol by User $k$ in P-NOMA and AP-NOMA methods and $P_{kn}$ is the magnitude assigned to the transmission of the $n$th symbol by User $k$ in T-NOMA method. In all methods, the transmit power constraint is $NP$ in $N$ symbol times.

\section{Achievable Rate-Region Analysis}\label{analysis}

This section analyzes the achievable rate-region of the proposed NOMA schemes, AP-NOMA and T-NOMA. For comparison purposes, we also present the achievable rate-regions of the synchronous P-NOMA.

\subsection{Achievable Rates for Conventional  P-NOMA}

We begin by applying the SIC detection at each user in the synchronous NOMA scheme. The
optimal detection sequence is $x_K, \cdots,x_2, x_1$ assuming $\frac{|h_1|
^2}{\sigma^2_n} > \cdots > \frac{|h_K|^2}{\sigma^2_n}$. In particular, User 1 first decodes $x_2, \cdots, x_K$ and subtracts
their components from the received signal $y_1$. Then, User 1 decodes $x_1$ without interference from other users. On the other hand,
User $K$ can directly decode $x_K$ while considering other users as noise.
Assuming successful decoding and no error propagation, the
achievable rate-region can be represented as:

\begin{flalign}
R_{P-NOMA}\triangleq \left\{\begin{matrix}
0\leq R_1 \leq \frac{1}{2}\log_2\left( 1+\frac{P_1|h_1|^2}{\sigma^2_n}\right)\\  
0\leq R_2 \leq \frac{1}{2}\log_2\left( 1+\frac{P_2|h_2|^2}{P_1|h_2|^2+\sigma^2_n}\right)\\  
\vdots\\  
0\leq R_K \leq \frac{1}{2}\log_2\left( 1+\frac{P_K|h_K|^2}{|h_K|^2\sum_{k=1}^{K-1}{P_k}+\sigma^2_n}\right)    
\end{matrix}\right.
\end{flalign}

\subsection{Achievable Rates for AP-NOMA}

In AP-NOMA, the SIC detection is the same as P-NOMA, however, the set of samples that is used to decode symbols of each user is different. In the synchronous case, there is only one set of samples at each user, namely, User $r$, with no ISI and can be written as $\boldsymbol{y_r^r}=h_r\sum_{k=1}^{K}{\boldsymbol{I_Nx_k}}+\boldsymbol{n_r^r}$. Based on the sufficient statistics derived in Eq. (\ref{suff}), in the asynchronous case, there are $K$ sets of samples at each user, each of them matched to the timing offset of one of the users represented as:
\begin{flalign}
\boldsymbol{y_j^r}=h_r\boldsymbol{}\sum_{\substack{j=1}}^{K}{\sqrt{P_k}\boldsymbol{R_{jk}x_k}}+\boldsymbol{n_j^r}, \ \ \ j=1,\cdots, K
\end{flalign}
Based on the assumption of $\frac{|h_1|
^2}{\sigma^2_n} > \cdots > \frac{|h_K|^2}{\sigma^2_n}$, User $r$ can decode signals of Users $\left\{r+1,  \cdots ,K\right\}$ using the sample sets of $\left\{\boldsymbol{y_{r+1}^r},  \cdots ,\boldsymbol{y_K^r}\right\}$ and subtract them from the corresponding sample set of $\boldsymbol{y_r^r}$. Note that based on the ordering of channel strengths, User $r$ only needs $K-r$ sets of samples. In particular, the strongest user needs all $K$ sets of samples and the weakest user only needs its own corresponding set of samples. The resulting set of samples after subtraction, at User $r$, is calculated as:
\begin{flalign}
\boldsymbol{\hat y_r^r}=h_r\sqrt{P_r}\boldsymbol{I_Nx_r}+h_r\boldsymbol{}\sum_{\substack{k=1}}^{r-1}{\sqrt{P_k}\boldsymbol{R_{rk}x_k}}+\boldsymbol{n_r^r}
\end{flalign}
Due to the Toeplitz structure of the sub-blocks $\boldsymbol{R_{lk}}$, the resulting sample at each time instant $i$ can be written as:
\begin{flalign}
\nonumber
\hat y_r^r[i]=h_r\sqrt{P_r}x_r[i]+h_r\sum_{\substack{k=1}}^{r-1}\sqrt{P_k}\sum_{j=-u}^{u}{g_{rk}(j)x_k[i-j]}
+n_r^r[i], \ \ \  \ i=u+1,\cdots,N-u
\end{flalign}
Treating the remaining interferers as noise will result in the following achievable rate for User $r$:
\begin{flalign}\label{aa}
0\leq R_r \leq \frac{1}{2}\log_2\left( 1+\frac{P_r|h_r|^2}{|h_r|^2\sum_{k=1}^{r-1}{G_{rk}P_k}+\sigma^2_n}\right) 
\end{flalign}
where $G_{rk}=\sum_{i=-u}^{u}{g^2_{rk}(i)}$. In general, the overall rate-region depends on how we assign the timing offsets to users with different channel strengths. For a specific assignment, namely $\psi$, the achievable rate-region, $R_{AP-NOMA}^{\psi}$ can be defined as: 

\begin{flalign}
R_{AP-NOMA}^{\psi}\triangleq\left\{\begin{matrix}
0\leq R_1 \leq \frac{1}{2}\log_2\left( 1+\frac{P_1|h_1|^2}{\sigma^2_n}\right)\\  
0\leq R_2 \leq \frac{1}{2}\log_2\left( 1+\frac{P_2|h_2|^2}{G_{21}P_1|h_2|^2+\sigma^2_n}\right)\\
\vdots\\  
0\leq R_K \leq \frac{1}{2}\log_2\left( 1+\frac{P_K|h_k|^2}{|h_K|^2\sum_{k=1}^{K-1}{G_{Kk}P_k}+\sigma^2_n}\right)    
\end{matrix}\right.
\end{flalign}

Note that there are $K!$ different assignments of time delays to users. Hence, the total rate-region is the convex hull of all possible assignments, i.e., 
\begin{flalign}
R_{AP-NOMA}\triangleq\bigcup_{\psi=1}^{K!} R_{AP-NOMA}^{\psi}
\end{flalign}

Using Lemma 1, it can be shown that $\sum_{-u}^{u}{g^2_{rk}(i)}<IUI(\tau)<IUI(0)=1$ where the last identity is valid for all pulse shapes that satisfy the Nyquist no-ISI condition, including, the rectangular, the sinc and the raised cosine pulse shapes. As a result, the rate-region for each assignment and thus the total rate-region for AP-NOMA is larger than that of the conventional P-NOMA. Note that for the 2-user scenario, the rate-region for both possible assignments is the same because $G_{21}=G_{12}$. However, this is not valid for more number of users unless the difference between time delays is equal which results in matrix $\boldsymbol{R}$ to be Hermitian Toeplitz. The numerical results are shown in Section \ref{simulation}.
\subsection{Achievable Rates for T-NOMA}

In this section, we derive the achievable rate-region for the T-NOMA method. As mentioned before, asynchrony provides additional degrees of freedom which will be exploited in T-NOMA by using precoding at the BS. T-NOMA method applies a simple precoding at the BS. In more details, after power assignment to users' intended symbols, they are precoded by a unitary matrix, i.e., $\boldsymbol{x_T}=\boldsymbol{U_TPx}$. Then, the received signal at User $r$ is calculated by:
\begin{flalign}
\boldsymbol{y^r}=h_r\boldsymbol{RU_TPx}+\boldsymbol{n^r}
\end{flalign}
where the power constraint is stated as $trace(\boldsymbol{RQ}) \leq NP$. The covariance matrix of the transmitted vector and noise vector are $\boldsymbol{Q}=\boldsymbol{U_T}\boldsymbol{P}^2\boldsymbol{U_T}^H$ and $\boldsymbol{Q_n}=\boldsymbol{R}\sigma^2_n$, respectively. 

To find the proper precoding matrix, let us consider the eigen-decomposition of matrix $\boldsymbol{R}$. Matrix $\boldsymbol{R}$ is a Hermitian matrix, thus its  eigen-decomposition can be written as:

\begin{flalign}
\nonumber
\boldsymbol{R}&=\boldsymbol{{U}_{R}} \left(\begin{matrix}
 \lambda_1&  0&  \cdots& 0\\ 
 0&  \lambda_2&  \cdots& 0\\ 
 \vdots& \ddots & \ddots & \vdots\\ 
 0& 0 & \cdots  &\lambda_{NK}
\end{matrix} \right)\boldsymbol{{U}_{R}}^H\\
&=\boldsymbol{{U}_{R}} \boldsymbol{\Lambda_R} \boldsymbol{{U}_{R}}^H
\end{flalign}

where $\lambda_{1}\geq \lambda_{2}\geq \cdots \geq \lambda_{NK}$ are the eigenvalues of matrix $\boldsymbol{R}$ and $\boldsymbol{{U}_R}$ is a Unitary matrix. Therefore, the received signal can be rewritten as:
\begin{flalign}
\boldsymbol{y^r}=h_r\boldsymbol{\boldsymbol{U}_{R} \boldsymbol{\Lambda_R} \boldsymbol{U}_{R}^HU_TPx}+\boldsymbol{n^r}
\end{flalign}

Because the time delays are known at the transmitter, the matrix $\boldsymbol{R}$ is known at the transmitter. Hence, the transmitted symbols can be precoded in the direction of the eigen-vectors of matrix $\boldsymbol{R}$. In addition, the sub-channels can be decomposed at each user by a post-processing matrix $\boldsymbol{U_U}$. By choosing $\boldsymbol{U_T}=\boldsymbol{U_R}$ and $\boldsymbol{U_U}=\boldsymbol{U_R}^H$, the received signal will be written as
\begin{flalign}
\boldsymbol{y^r}=h_r\boldsymbol{\Lambda_R}\boldsymbol{Px}+\boldsymbol{\hat n^r}
\end{flalign}
where $\boldsymbol{\hat n^r}=\boldsymbol{U}^H\boldsymbol{n^r}$ is a white Gaussian noise with covariance matrix of $\boldsymbol{Q_{\hat n}}=\boldsymbol{\Lambda_R}\sigma^2_n$. By using proper precoding and post processing at the destination, the channel has turned into $NK$ independent sub-channels. In other words, we have used the available degrees of freedoms offered by asynchrony to decompose the transmitted symbols and eliminate interference.  

As we will see later, the way of assigning different symbols to different sub-channels will not change the final result, thus, for notational simplicity, we denote the eigenvalue corresponding to $x_k[n]$ as $\lambda_{kn}$. Due to the channel decomposition, there is no interference and the achievable rate for each user is sum of the achievable rates at the corresponding sub-channels: 
\begin{flalign}
\nonumber
R_{r}=\frac{1}{2N}\sum_{i=1}^{N}{log_2\left(1+\frac{P_{ri}\lambda_{ri}|h_r|^2}{\sigma^2_n}\right)}
\end{flalign}
 
The precoding at the BS also affects the power constraint. The alignment of transmit covariance matrix in the direction of the eigen-vectors of matrix $\boldsymbol{R}$ changes the product of $\boldsymbol{RQ}$ to:
\begin{flalign}
\boldsymbol{RQ}=\boldsymbol{U}_{R} \left(\begin{matrix}
 P_{11}\lambda_{11}&  0&  \cdots& 0\\ 
 0&  P_{12}\lambda_{12}&  \cdots& 0\\ 
 \vdots& \ddots & \ddots & \vdots\\ 
 0& 0 & \cdots  &P_{KN}\lambda_{KN}
\end{matrix} \right)\boldsymbol{U}_{R}^H
\end{flalign}
As a result, the power constraint will be:
\begin{flalign}
\sum_{k=1}^{K}\sum_{i=1}^{N}{P_{ki}\lambda_{ki}}\leq NP
\end{flalign}
\begin{lemma}
The achievable rate-region of T-NOMA method can be described as follows:
\end{lemma}
\begin{flalign}
\nonumber
R_{r}&=\frac{1}{2}log_2\left(1+\frac{P_r|h_r|^2}{\sigma^2_n}\right)\\
&s.t. \sum_{r=1}^{K}{P_r}\leq P
\end{flalign}
\begin{IEEEproof}
The proof is presented in Appendix \ref{appendix3}.
\end{IEEEproof}
Note that the achievable rate-region of T-NOMA method is independent of the pulse shape and the timing offsets. The only requirement to achieve this rate-region is the matrix $\boldsymbol{R}$ to be full rank which is satisfied as long as the pulse shapes are time limited and the timed delays are distinct. Also note that, using relatively band-limited signals like truncated sinc pulse shapes and truncated Raised Cosine pulse shapes will prevent the spectrum broadening caused by precoding \cite{kim2013faster}. Besides the individual performance of each user, the other significant criterion in a network is the amount of cumulative rate of all users which can be achieved at a same time termed as sum-rate. In the next section, we compare the sum-rate provided by P-NOMA, AP-NOMA and T-NOMA methods. We will show that for the P-NOMA method, achieving the maximum available sum-rate is equivalent to violating the fairness; however, AP-NOMA and T-NOMA methods can achieve the maximum sum-rate while maintaining fairness among users. 
\subsection{Sum-Rate Analysis}
Although analyzing the sum-rate in the general case of $K$ users is cumbersome, considering the case of $K=2$ will shed some light on behavior of sum-rate and its optimum points. In the next lemma, we compare the sum-rate results for P-NOMA, AP-NOMA and T-NOMA and also their corresponding fairness. 
\begin{lemma}{\label{a}}
Denoting $R^*$ as the maximum of sum-rate, i.e., $R=R_1+R_2$, we will have the following results:
\begin{itemize}
\item Sum-Rate Comparison:
\begin{flalign}
R^*_{P-NOMA}<R^*_{AP-NOMA}<R^*_{T-NOMA}
\end{flalign}
\item Fairness:\\
In order to achieve the max sum-rate, one of the users is assigned zero power if:
 \begin{flalign}
\left\{\begin{matrix}
 |\sigma_1-\sigma_2|>0 \ \ & P-NOMA\\ 
 |\sigma_1- \sigma_2|>(1-g)P & AP-NOMA\\
 |\sigma_1- \sigma_2|> P & T-NOMA
\end{matrix}\right. 
\end{flalign}
\end{itemize}
where $\sigma_i=\frac{\sigma^2_n}{|h_i|^2}$ and $g$ is equal to $G_{12}(=G_{21})$ defined in Eq. (\ref{aa}). 
\end{lemma}
\begin{IEEEproof}
The proof is presented in Appendix \ref{appendix4}.
\end{IEEEproof}
The results in the first part of Lemma \ref{a} indicates that the T-NOMA method provides the largest sum-rate. The second part explains that in order to achieve the maximum sum-rate, P-NOMA almost always (i.e., for $\sigma_1\neq \sigma_2$) assigns zero power to one of the users, thus, achieving the maximum sum-rate with fairness is not possible using the P-NOMA method. However, AP-NOMA and T-NOMA methods assign zero power to one of the users only when the difference between $\sigma_1$ and $\sigma_2$ is greater than $(1-g)P$ and $P$, respectively. In other words, unlike the P-NOMA method, AP-NOMA and T-NOMA can achieve the maximum of the sum-rate while assigning non-zero power to both users. Numerical results are provided in the next section.
\section{Numerical Results}\label{simulation}
In this section, we present numerical results to show the effectiveness of using asynchrony in providing a larger rate-region. In particular, we show that AP-NOMA outperforms P-NOMA with a slight change of adding timing offset among transmitted symbols. However, T-NOMA which uses the degrees of freedom provided by time asynchrony results in the best performance. We assume that a transmit power of $P=10$ is available at the base station which is serving two users. We consider
the typical pulse shaping function in the literature, i.e., Rectangular pulse shape (Rect.), and a more practical pulse shaping function, i.e., Root Raised Cosine (R.R.C.). Theoretically, R.R.C. pulse shaping is unlimited in time, however, it is truncated in practice and we have adopted the truncated version with 4 side lobes. The symbol duration $T$ is normalized to be 1, $\tau_1=0$ and $\tau_2\in [0, 0.5]$ due to the symmetry.

\begin{figure}[h]
    \centering
    \includegraphics[width=3.3in]{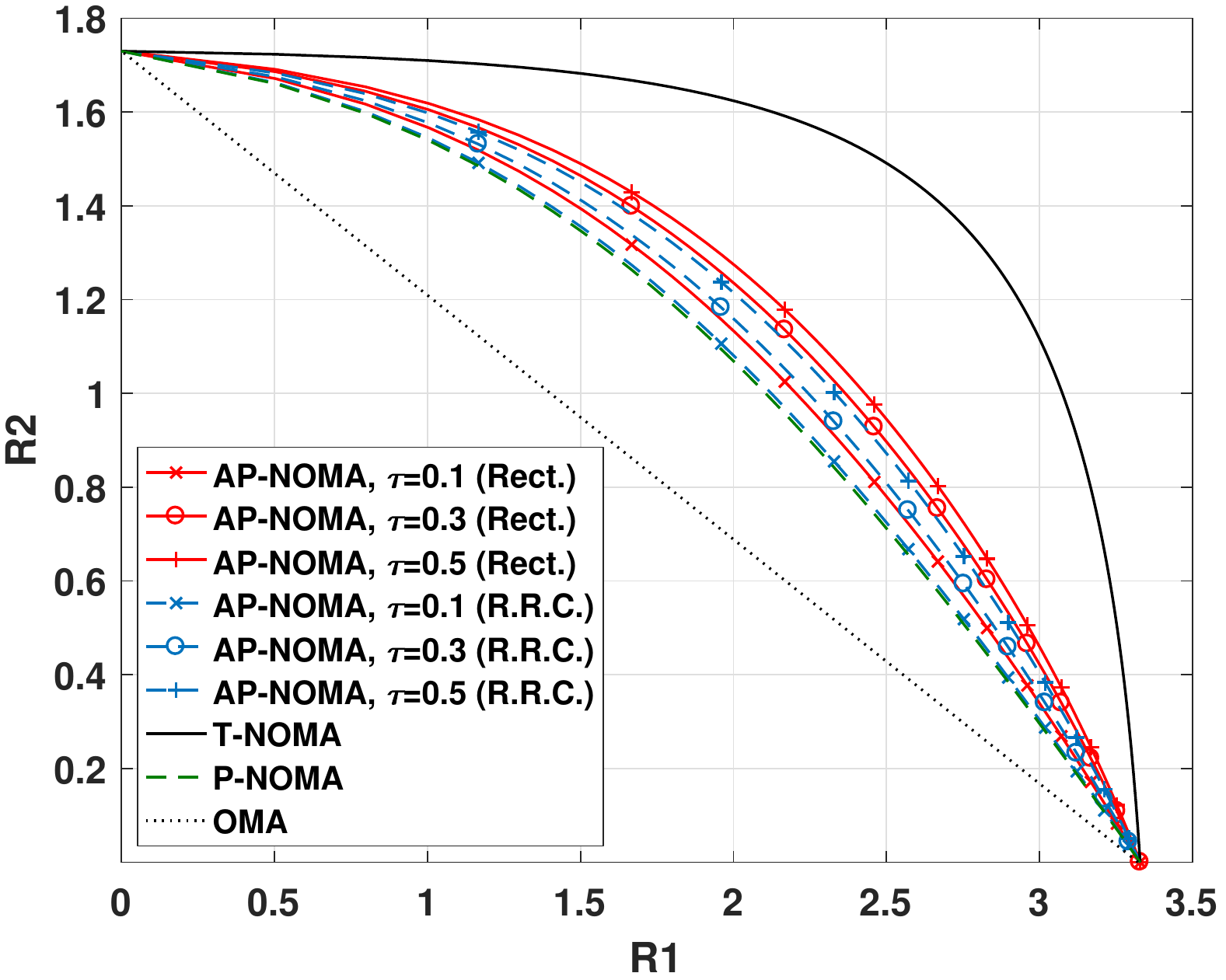}
    \caption{Achievable rate-regions of AP-NOMA and T-NOMA for different symbol offsets and a Gaussian channel.}
    \label{compare_delay}
\end{figure}

We first consider the Gaussian channel where the channel
coefficients are determined by ${|h_1|^2} / {\sigma^2_n}= 10$ and ${|h_2|^2} / {\sigma^2_n}=1$.
In Fig. \ref{compare_delay}, we show the achievable rate-regions of AP-NOMA and T-NOMA with different symbol offsets. 

\begin{figure}[h]
    \centering
    \includegraphics[width=3.3in]{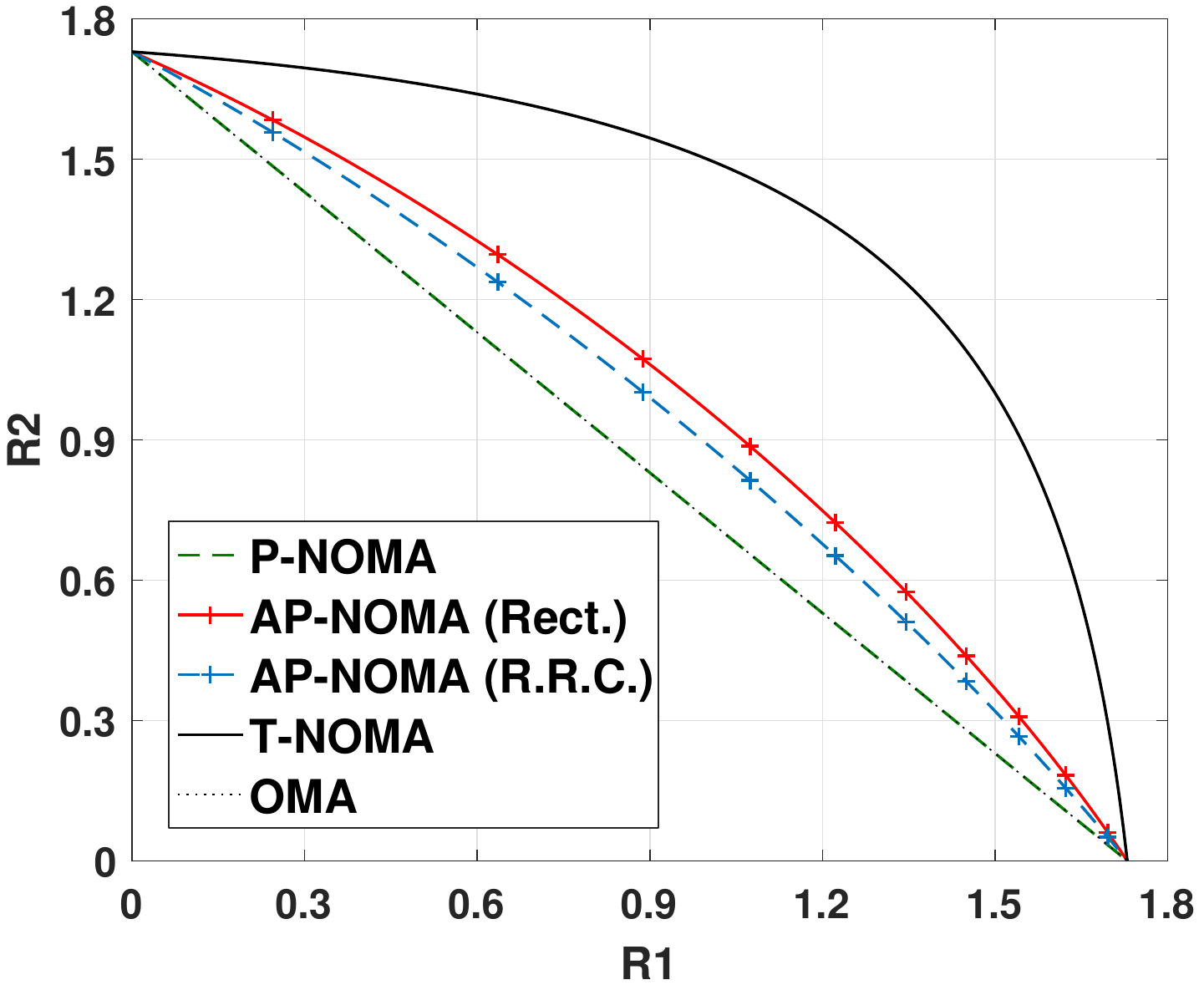}
    \caption{The maximum achievable rate-regions of three schemes: P-NOMA, AP-NOMA and T-NOMA in Gaussian channels with ${|h_1|^2} / {\sigma^2_n}= 1$ and ${|h_2|^2} / {\sigma^2_n}=1$.}
    \label{compare_1_1}
\end{figure}

In the AP-NOMA method, the choice of pulse shaping affects the amount of reduction in IUI and thus the overall performance. Rect. pulse shaping provides slightly better performance compared with R.R.C. For both Rect. and R.R.C. pulse shapes, increasing the timing offset will improve the performance and $\tau_2=0.5$ results in the best performance. On the other hand, the performance of the T-NOMA method, which exploits the degrees of freedom available in the system, is independent of the pulse shape and time delays as long as matrix $\boldsymbol{R}$ is full rank. Fig. \ref{compare_delay} shows that both T-NOMA and AP-NOMA outperform the conventional P-NOMA. In Fig. \ref{compare_delay}, there is sufficient discrepancy between channel coefficients of the two users to be exploited by P-NOMA and AP-NOMA methods. However, in Fig. \ref{compare_1_1}, the quality of channel coefficients are assumed to be the same, i.e., ${|h_1|^2} / {\sigma^2_n}= 1$ and ${|h_2|^2} / {\sigma^2_n}=1$, thus the P-NOMA performance coincides with that of the OMA systems like TDMA. In such a case, the AP-NOMA method provides slightly better performance; however, T-NOMA significantly improves the performance showing the capability of this method even without power discrepancy.  

\begin{figure}[h]
    \centering
    \includegraphics[width=3.3in]{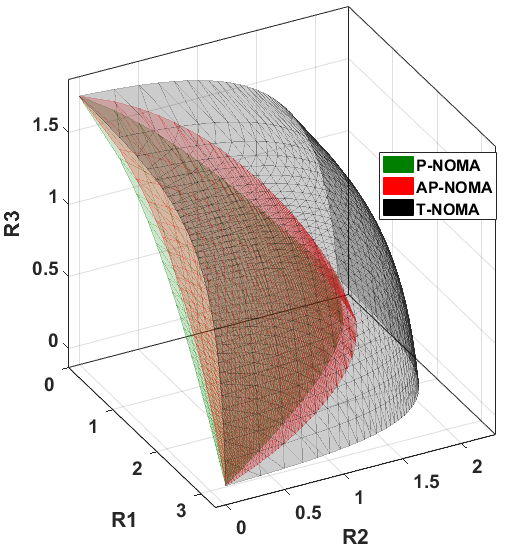}
    \caption{The maximum achievable rate-regions of three schemes: P-NOMA, AP-NOMA and T-NOMA in Gaussian channels with ${|h_1|^2} / {\sigma^2_n}= 10$, ${|h_2|^2} / {\sigma^2_n}=2$ and ${|h_3|^2} / {\sigma^2_n}= 1$.}
    \label{3d}
\end{figure}

The rate-region for 3 users with rectangular pulse shape and $\boldsymbol{\tau}=[0,0.3,0.7]$ is shown in Fig. \ref{3d}. The rate-regions for P-NOMA and T-NOMA are calculated similar to those of the 2-user scenario. However, there are two main differences in calculating the rate-region of AP-NOMA. First, there are 3!=6 different assignments of time delays to users, and the total region is found by taking the convex hull of all possible assignments. The other difference is that when one of the power assignments is equal to zero, the remaining time delays for the other two users need to be updated. In more details:
\begin{flalign}
R_{AP-NOMA}=\left\{\begin{matrix}
R_{AP-NOMA}[0,0.3,0.7] \ \ \ \   P_1\neq 0, P_2\neq 0,P_3\neq 0\\  R_{AP-NOMA}[0,0.4] \ \ \ \ \  P_1= 0, P_2\neq 0,P_3\neq 0\\
R_{AP-NOMA}[0,0.7] \ \ \ \ \ P_1\neq 0, P_2=0,P_3\neq 0\\ 
R_{AP-NOMA}[0,0.3] \ \ \ \ \ P_1\neq 0, P_2\neq0,P_3= 0\\  
\end{matrix}\right.
\end{flalign}

\begin{figure}[h]
\centering
\subfloat[$R_1$=1]{\includegraphics[width=2in]{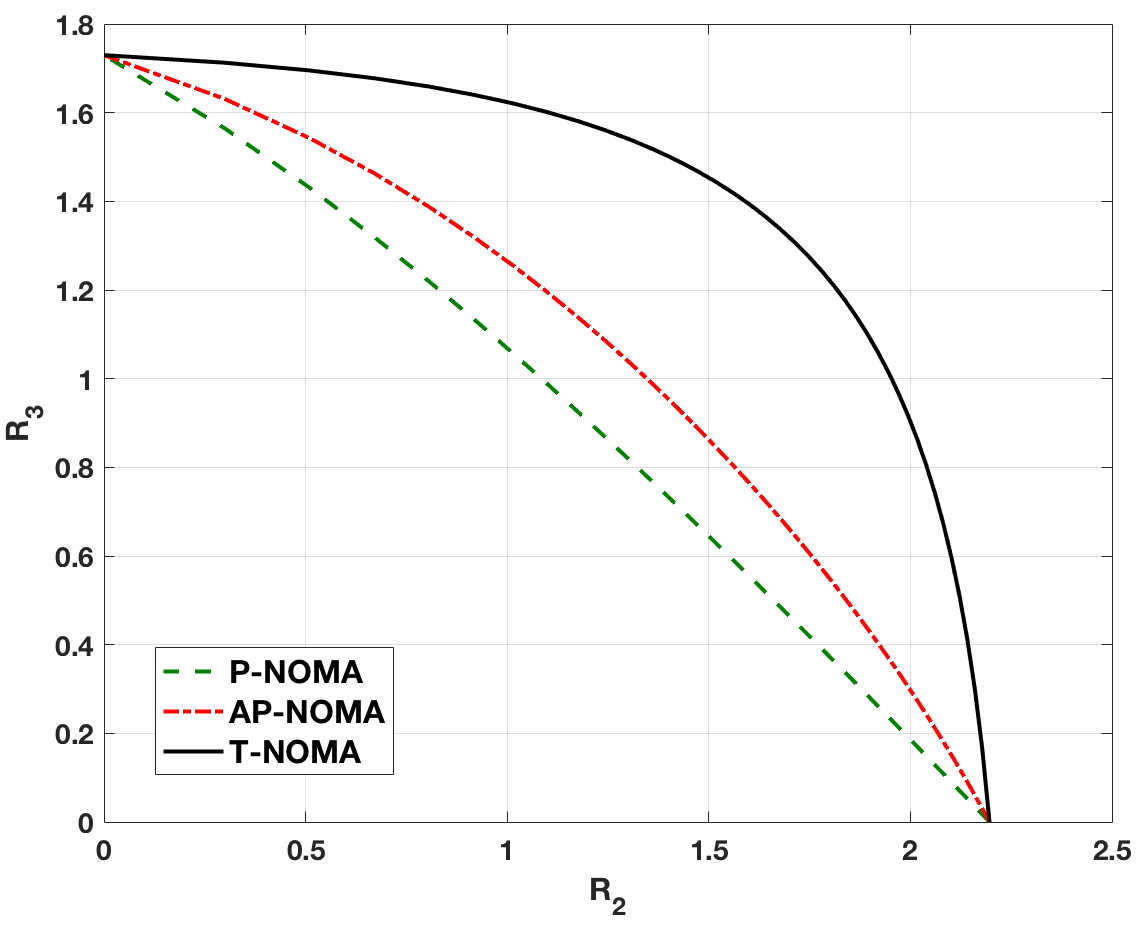}
\label{cut_a}}
\hfil
\subfloat[$R_2$=1]{\includegraphics[width=2in]{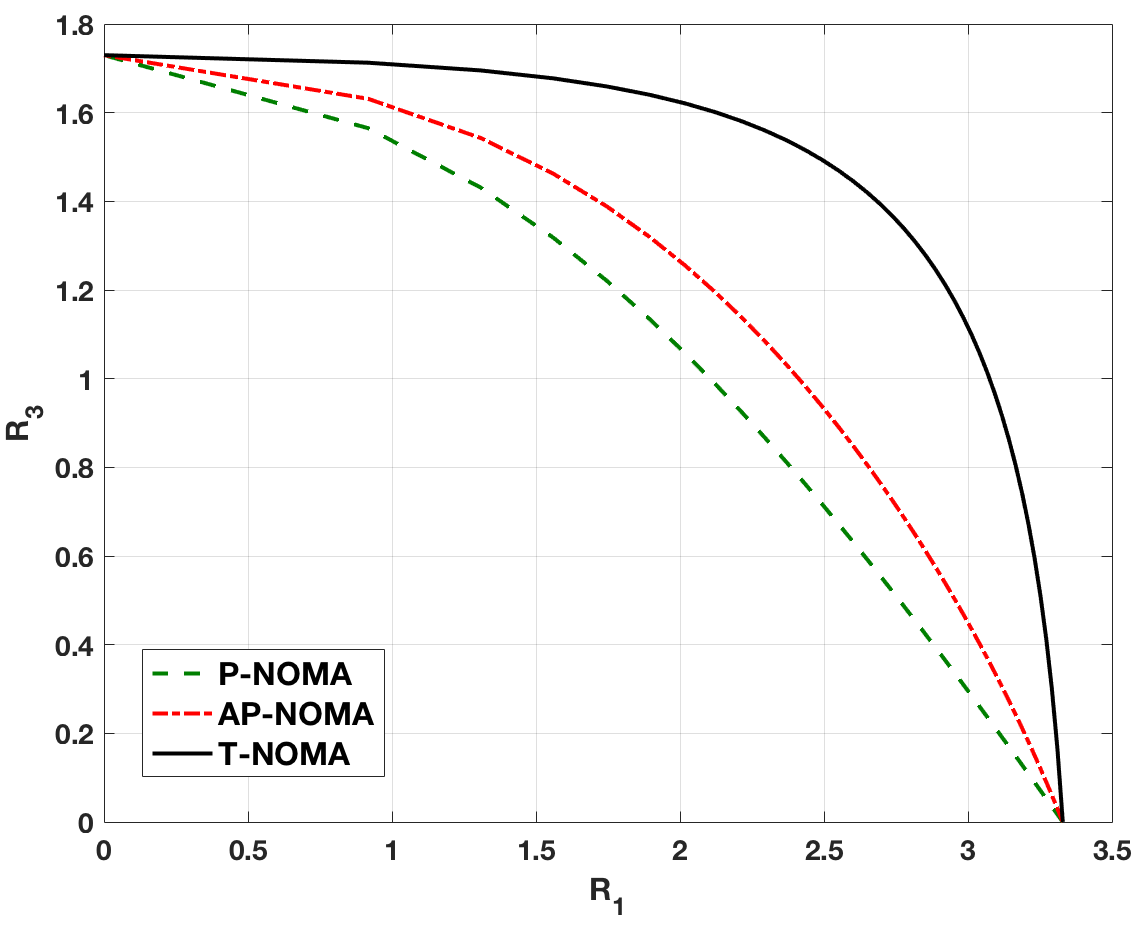}
\label{cut_b}}
\hfil
\subfloat[$R_3$=1]{\includegraphics[width=2in]{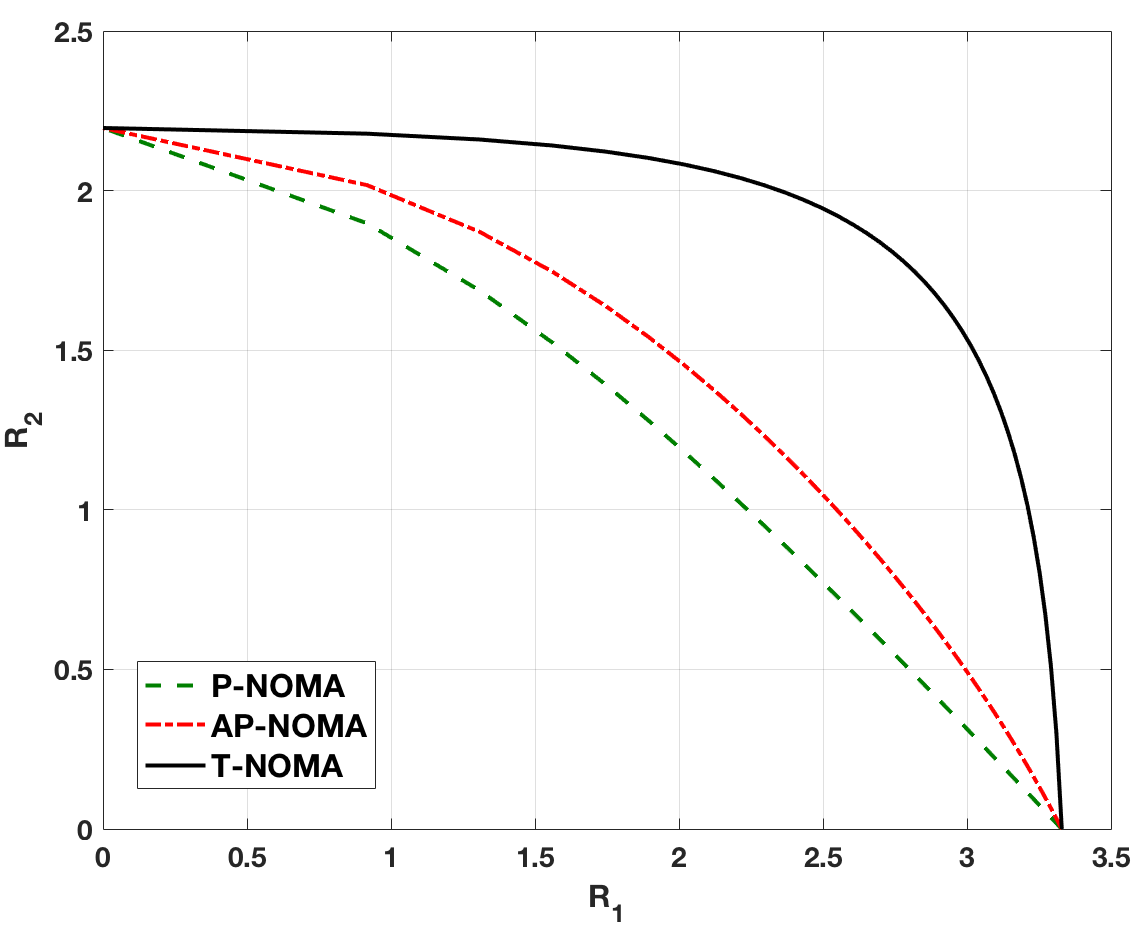}
\label{cut_c}}
\caption{Different 2D cuts of the 3 dimensional rate-region in Fig. \ref{3d}}
\end{figure}

To have a better understanding of the 3 dimensional rate-region provided in Fig. (\ref{3d}), we show different two-dimensional cuts when $R_1=1$, $R_2=1$ and $R_3=1$ in Figs. (\ref{cut_a}), (\ref{cut_b}) and (\ref{cut_c}), respectively.

\begin{figure}[h]
    \centering
    \includegraphics[width=3.3in]{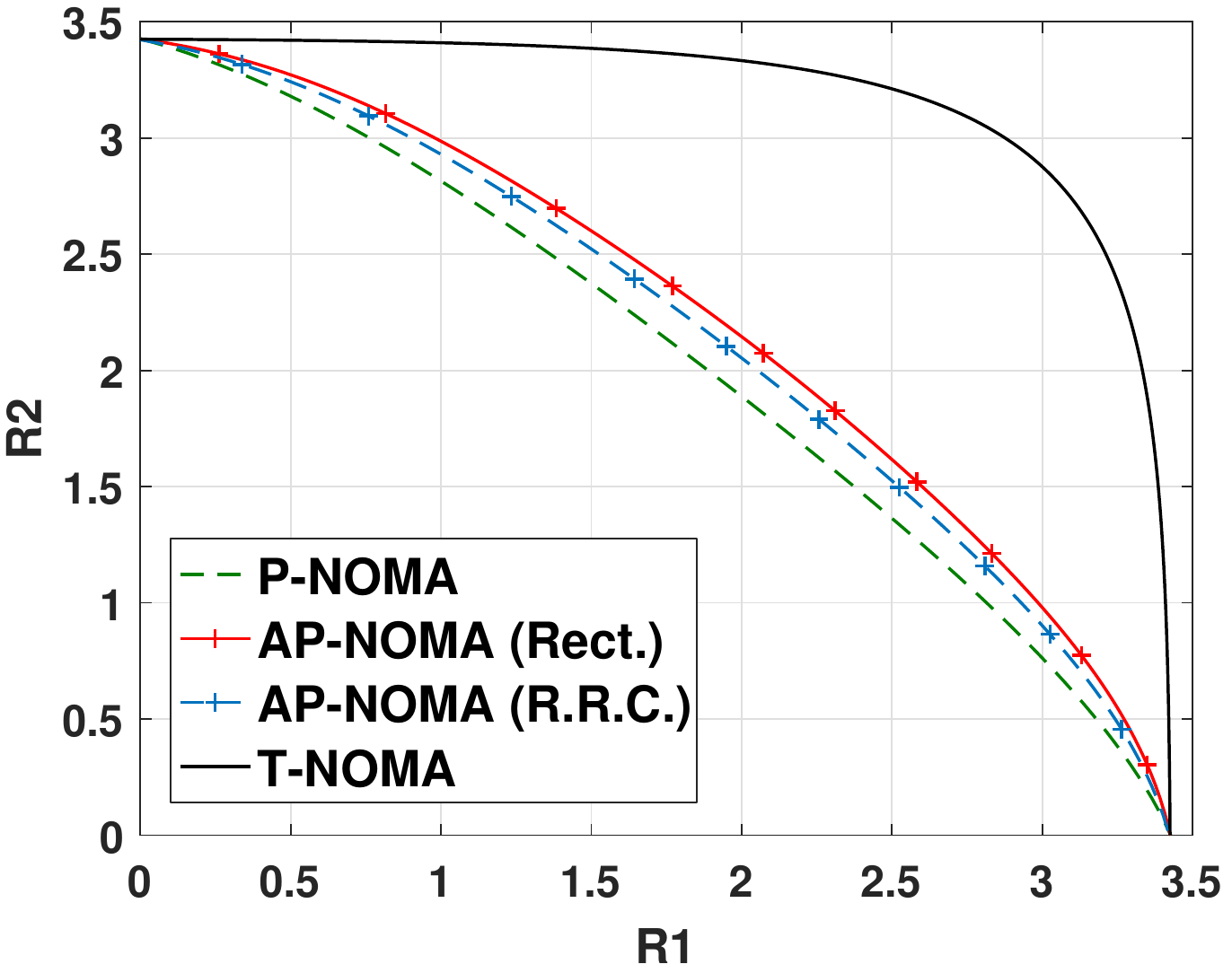}
    \caption{The maximum achievable rate-regions of three schemes: P-NOMA, AP-NOMA and T-NOMA in Rayleigh fading channels.}
    \label{compare_fading}
\end{figure}

Next, we consider the Rayleigh block-fading channel where
the channel coefficients are independent Rayleigh distribution
with unit variance, and noise variance is set to 0.1. The ergodic rate is averaged over $10^5$ different realizations of the channel. In Fig. \ref{compare_fading}, we show the achievable rate-regions of P-NOMA, AP-NOMA and T-NOMA in Rayleigh
block-fading channels. As was the case in Gaussian channels, the achievable rate-region of P-NOMA is improved by adding asynchrony. In addition, T-NOMA provides a large improvement compared with other schemes.

\begin{figure}[h]
    \centering
    \includegraphics[width=3.3in]{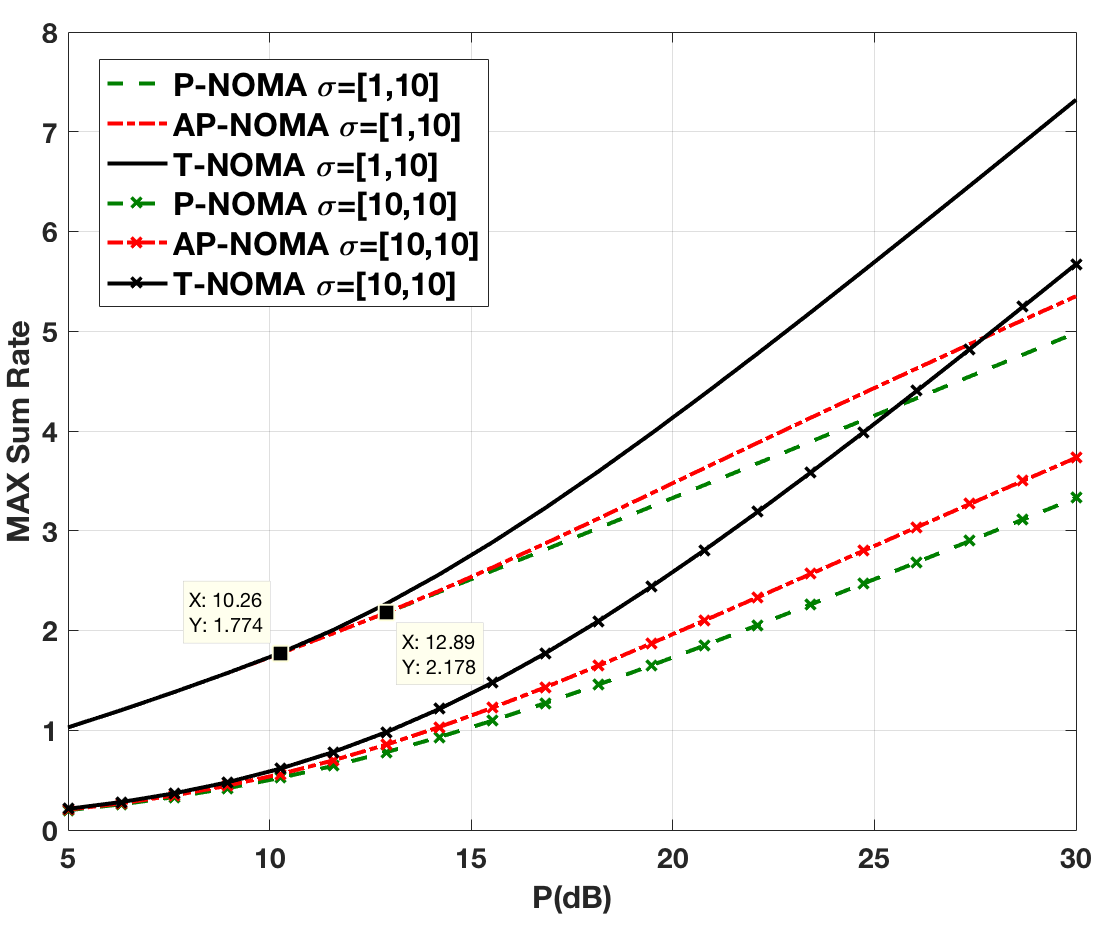}
    \caption{The maximum sum-rate with respect to available transmit power for two cases: different channel qualities and same channel qualities.}
    \label{sumrate}
\end{figure}

In Fig. \ref{sumrate}, the maximum sum-rate with respect to the available transmit power is presented. In the case where channels have the same qualities (i.e., $\sigma_1=\sigma_2$), both AP-NOMA and T-NOMA methods are strictly better than P-NOMA. For the case of different channel qualities (i.e., $\sigma_1\neq\sigma_2$), AP-NOMA outperforms the P-NOMA method when $P\geq \frac{\sigma_2-\sigma_1}{g}\approx 13 dB$ and T-NOMA outperforms the P-NOMA method when $P\geq \frac{\sigma_2-\sigma_1}{g}\approx 10 dB$. When $\sigma_1=0.1$ and $\sigma_2=1$, P-NOMA assigns all the power to the first user, however, AP-NOMA and T-NOMA assign non-zero powers to both users when  $P\geq 13$ and $P\geq 10$, respectively. For example when $P=45$ dB power is available, AP-NOMA assigns $P_1=20.3, P_2=24.7$ and T-NOMA assigns $P_1=22.8, P_2=22.2$ to the first and second users, respectively. Therefore, not only do AP-NOMA and T-NOMA outperform P-NOMA but also they maintain the fairness among the users. 

\section{Conclusion}\label{conclusion}
In this work, we propose novel symbol-asynchronous downlink NOMA schemes. In contrast to the conventional P-NOMA, we propose to intentionally add timing offsets among superimposed symbols. The receiver architecture in AP-NOMA includes oversampling and a SIC scheme similar to the P-NOMA, however, asynchrony reduces IUI and improves the overall performance. T-NOMA exploits the degrees of freedom introduced by time asynchrony, using novel precoding and simple post processing at users. In other words, T-NOMA decomposes the channel into independent sub-channels and eliminates the interference. Our analysis shows that both AP-NOMA and T-NOMA methods can improve the achievable rate-regions. In addition, we showed that the proposed methods provides higher sum-rate while maintaining the fairness among the users. 

\appendices

\section{Proof of Lemma 1}\label{appendix1}
Denoting g(t) as a pulse shape with real spectrum, we show that $IUI(\tau)$ calculated as:
\begin{flalign}
IUI(\tau)=\sum_{i=-\infty}^{\infty}{|g(\tau+iT)|^2}
\end{flalign}
is maximized at $\tau=0$. In other words, $IUI(\tau)<IUI(0), \ \ \tau\neq 0$.
Assume that $G(f)$ is the Fourier transform of the pulse shape $g(t)$. Then, the Fourier transform of the shifted version of $g(t)$, i.e., $g(\tau+t)$, will be $G(f)e^{j2\pi f \tau}$. The DTFT of the samples of $g(\tau+t)$, i.e., $g(\tau+iT), i\in Z$ can be expressed as:
\begin{flalign}
G'_{\tau}(f)=\sum_{i=-\infty}^{\infty}{G(f+i/T)e^{j2\pi (f+i/T) \tau}}
\end{flalign}
Note that $G'_{\tau}(F)$ is periodic with period of $1/T$. Based on the Parseval's theorem, the IUI energy, i.e., $IUI(\tau)=\sum_{i=-\infty}^{\infty}{|g(\tau+iT)|^2}$, will be equal to:
\begin{flalign}
IUI(\tau)=\int_{-1/2T}^{1/2T}{|G'_{\tau}(f)|^2df}
\end{flalign}
Then, with the assumption of having real spectrum, we will have:
\begin{flalign}
\nonumber
IUI(\tau)&=\int_{-1/2T}^{1/2T}{\left|\sum_{i=-\infty}^{\infty}{G(f+i/T)e^{j2\pi (f+i/T) \tau}}\right|^2df}\\
&\leq \int_{-1/2T}^{1/2T}{\left|\sum_{i=-\infty}^{\infty}{G(f+i/T)}\right|^2df}=IUI(0)
\end{flalign}
which concludes the proof.
\section{Proof of Lemma 2}\label{appendix2}

The power of the asynchronous signal can be written as:
\begin{flalign}
P_{asynch}&=E\left[\int_{-\infty}^{\infty}x_{asynch}(t)x^*_{asynch}(t)dt \right]\\
\nonumber
&=E\left[\int_{-\infty}^{\infty} \left(\sum _{k=1}^{K}x_k(t-\tau_k)\right)\left(\sum _{k=1}^{K}x_k(t-\tau_k)\right)^*dt \right]
\end{flalign}
Then, we have
\begin{flalign}
\nonumber
P_{asynch}&=
E\left[\int_{-\infty}^{\infty} \left(\sum_{k=1}^K\sum_{n=1}^{N}{x_k[n]p(t-(n-1)T-\tau_k)}\right)\left(\sum_{k=1}^K\sum_{n=1}^{N}{x_k[n]p(t-(n-1)T-\tau_k)}\right)^*dt \right]\\
\nonumber
&=\sum_{k_1=1}^K\sum_{k_2=1}^{K}\sum_{n_1=1}^N\sum_{n_2=1}^{N}E\left[x_{k_1}[n_1]x_{k_2}[n_2]\right] \int_{-\infty}^{\infty} p(t-(n_1-1)T-\tau_{k_1})p(t-(n_2-1)T-\tau_{k_2})dt \\
&=\sum_{k_1=1}^K\sum_{k_2=1}^{K}trace(\boldsymbol{R_{k_1k_2}}COV[\boldsymbol{x_{k_1}},\boldsymbol{x_{k_2}}])=trace(\boldsymbol{R}\  COV[\boldsymbol{x}])
\label{big}
\end{flalign}
which concludes the proof.
\section{Proof of Lemma 3}\label{appendix3}
The achievable rate for each user is the sum of the achievable rates for the corresponding sub-channels: 
\begin{flalign}
R_{r}=\frac{1}{2N}\sum_{i=1}^{N}{log_2\left(1+\frac{P_{ri}\lambda_{ri}|h_r|^2}{\sigma^2_n}\right)}
\end{flalign}
with the power constraint of:
\begin{flalign}
\sum_{k=1}^{K}\sum_{i=1}^{N}{P_{ki}\lambda_{ki}}\leq NP
\end{flalign}
Denoting $p_{ri}$ as $\frac{P_{ri}\lambda_{ri}}{N}$ and $P_r$ as $\sum_{i=1}^{N}{p_{ri}}$, the above optimization problem can be rewritten in $K$ simpler problems as:
\begin{flalign}
\nonumber
R_{r}&=\frac{1}{2N}\sum_{i=1}^{N}{log_2\left(1+\frac{p_{ri}N|h_r|^2}{\sigma^2_n}\right)}\\
&s.t.\ \  \sum_{i=1}^{N}{p_{ri}}=P_r
\end{flalign}
The sum of the power constraints for the sub-problems should add up to $P$, i.e., $\sum_{r=1}^{K}{P_r}\leq P$.
It can be easily shown that the power assignment that maximizes $R_r$ is such that:
\begin{flalign}
p_{ri}=\cdots=p_{rN}=P_r/N
\end{flalign}
Therefore, by simple substitution, we can conclude that the achievable rate for each user is:
\begin{flalign}
\nonumber
R_{r}=\frac{1}{2}{log_2\left(1+\frac{P_{r}|h_r|^2}{\sigma^2_n}\right)}
\end{flalign}
such that $\sum_{r=1}^{K}{P_r}\leq P$. Note that the different assignments of sub-channels to users only change the power assignment, otherwise, the final result remains the same. 

\section{Proof of Lemma 4}\label{appendix4}
Using (\ref{aa}) for two users, the achievable sum-rate by AP-NOMA can be calculated as: 
\begin{flalign}\label{sum}
R=\frac{1}{2}\log_2\left( 1+\frac{P_1}{\sigma_1}\right)+\frac{1}{2}\log_2\left( 1+\frac{P_2}{gP_1+\sigma_2}\right)
\end{flalign}
where $\sigma_i=\frac{\sigma^2_n}{|h_i|^2}$ (assuming $\sigma_2\geq\sigma_1$ without loss of generality). By inserting $P_2=P-P_1$, taking the derivative with respect to $P_1$, making it equal to zero, and discarding the non-relevant terms, we will have:
\begin{flalign}\label{qw}
g(g-1)(P^*_1)^2+2(g-1)\sigma_2(P^*_1)+P(\sigma_2-g\sigma_1)+\sigma_2(\sigma_2-\sigma_1)=0
\end{flalign}
where $P_1^*$ is the optimal power allocated to the first user. By inserting $g=1$, Eq. (\ref{qw}) results in  $(P+\sigma_2)(\sigma_2-\sigma_1)$ for P-NOMA. Thus, for P-NOMA, if $\sigma_2=\sigma_1$, then the derivative is always equal to zero, meaning that the sum-rate is a constant value. In fact, if we put $\sigma_1=\sigma_2=\sigma$ and $g=1$, Eq. (\ref{sum}) simplifies to:
\begin{flalign}\label{sum2}
R=\frac{1}{2}\log_2\left( 1+\frac{P}{\sigma}\right)
\end{flalign}
In other words, when the channels have the same quality, the sum-rate of P-NOMA is always fixed, independent of the power assignment. However, if $\sigma_2>\sigma_1$, the derivative with respect to $P_1$ is always positive, implying that the sum-rate is strictly increasing in $P_1$. Thus, assigning $P_1=P$ results in the maximum sum-rate of P-NOMA which will be equal to $R^*=\frac{1}{2}\log_2\left( 1+\frac{P}{\sigma_1}\right)$. Note that for $\sigma_1>\sigma_2$, all the power will be assigned to $P_2$ and the sum-rate will be $R^*=\frac{1}{2}\log_2\left( 1+\frac{P}{\sigma_2}\right)$. 

Unlike P-NOMA where the derivative is always positive (or zero when $\sigma_1=\sigma_2$), for AP-NOMA, Eq. (\ref{qw}) has two roots which one of them is infeasible and the other one is:
\begin{flalign}\label{P1}
P_1^*=-\frac{\sigma_2}{g}+\frac{\sqrt{\sigma_2^2+\frac{g}{1-g}A}}{g}
\end{flalign}
where $A=P(\sigma_2-g\sigma_1)+\sigma_2(\sigma_2-\sigma_1)$. If $P^*_1\geq P$, i.e., $P\leq \frac{\sigma_2-\sigma_1}{1-g}$, then all the power is assigned to User 1, i.e., $P^*_1=P$. It can be concluded that as long as $|\sigma_1- \sigma_2|<(1-g)P$, then non-zero powers will be assigned to both users. 

To prove that $R^*_{P-NOMA}<R^*_{AP-NOMA}$, we first find the values of $P_1^*$ such that:
\begin{flalign}\label{badbakht}
\frac{1}{2}\log_2\left( 1+\frac{P^*_1}{\sigma_1}\right)+\frac{1}{2}\log_2\left( 1+\frac{P-P_1^*}{gP^*_1+\sigma_2}\right)>\frac{1}{2}\log_2\left( 1+\frac{P}{\sigma_1}\right)
\end{flalign}
After some calculations, it can be found that Inequality (\ref{badbakht}) is satisfied if $P_1^*>\frac{\sigma_2-\sigma_1}{1-g}$. On the other hand, considering the assumption of $P> \frac{\sigma_2-\sigma_1}{1-g}$ in Eq. (\ref{P1}) results in $P_1^*>\frac{\sigma_2-\sigma_1}{1-g}$. Hence, the we can conclude that $R^*_{P-NOMA}<R^*_{AP-NOMA}$ for $P> \frac{\sigma_2-\sigma_1}{1-g}$. In summary, if $P\leq \frac{|\sigma_2-\sigma_1|}{1-g}$, then $R^*_{AP-NOMA}=R^*_{P-NOMA}=\frac{1}{2}\log_2\left( 1+\frac{P}{min\{\sigma_1,\sigma_2\}}\right)$ which is achieved by assigning the total available power to the stronger user, otherwise, AP-NOMA assigns non-zero power to both users and $R^*_{P-NOMA}<R^*_{AP-NOMA}$. Note that, when $\sigma_2=\sigma_1$, the sum-rate achieved by AP-NOMA is strictly greater than the one achieved by P-NOMA and both users will be assigned non-zero powers. 

The sum-rate for T-NOMA can be calculated as:
\begin{flalign}
R=\frac{1}{2}log_2\left(1+\frac{P_1}{\sigma_1}\right)+\frac{1}{2}log_2\left(1+\frac{P_2}{\sigma_2}\right)
\end{flalign}
By inserting $P_2=P-P_1$, taking the derivative with respect to $P_1$, making it equal to zero we can find the optimal $P_1$ as:
\begin{flalign}
P_1^*=\frac{P+\sigma_2-\sigma_1}{2}
\end{flalign}
Therefore, if $\sigma_2-\sigma_1>P$, then all available power is assigned to User 1, i.e., $P^*_1=P$. Similarly, it can be shown that if $P\leq {|\sigma_2-\sigma_1|}$, then $R^*_{T-NOMA}=R^*_{P-NOMA}=\frac{1}{2}\log_2\left( 1+\frac{P}{min\{\sigma_1,\sigma_2\}}\right)$ which is achieved by assigning the total available power to one of the users, otherwise, T-NOMA assigns non-zero power to both users and $R^*_{P-NOMA}<R^*_{AP-NOMA}<R^*_{T-NOMA}$. The superiority of the maximum sum-rate achieved by T-NOMA can be easily verified by the fact that every optimal pair of $(P^*_1,P^*_2)$ for P-NOMA or AP-NOMA will result in a higher sum-rate for T-NOMA. 

\bibliographystyle{IEEEtran}
\bibliography{thesis}

\addtolength{\textheight}{-12cm}   




\end{document}